\documentclass{aa}
\usepackage{natbib}
\usepackage{graphicx}
\usepackage{amssymb}
\usepackage{color}

\newcommand{\bl}[1]{\mbox{\boldmath$ #1 $}}
\newcommand{\msun}{\mbox{M$_\odot$}}
\newcommand{\spitzer}{\textit{Spitzer}}
\newcommand{\lint}{\mbox{L$_{\rm int}$}}
\newcommand{\lsun}{\mbox{L$_\odot$}}
\newcommand{\rsun}{\mbox{R$_\odot$}}

\begin{document}

\title{The nature of very low luminosity objects (VeLLOs)}

\author{Eduard I. Vorobyov \inst{1,2}, Vardan Elbakyan \inst{2}, Michael M. Dunham \inst{3}, and  Manuel Guedel \inst{1}}
\authorrunning{Vorobyov et al.}

\institute{University of Vienna, Department of Astrophysics,  Vienna, 1180, Austria; 
\email{eduard.vorobiev@univie.ac.at}
\and
Research Institute of Physics, Southern Federal University, Stachki Ave. 194, Rostov-on-Don, 
344090 Russia; 
\and
 Department of Physics, State University of New York at Fredonia, 280 Central Avenue, Fredonia, NY 14063, USA
}

\abstract 
{}
 {The nature of very low luminosity objects with the internal luminosity  
 $L_{\rm obj}\le 0.1~L_\odot$ is investigated by means of numerical modeling
 coupling the core collapse simulations with the stellar
 evolution calculations.}
 {The gravitational collapse of a large sample of model cores in the mass range $0.1-2.0~M_\odot$ 
 is investigated. Numerical simulations 
 were started at the pre-stellar phase and terminated at the end of the embedded
phase when 90\% of the initial core mass had been accreted
onto the forming protostar plus disk system.  The disk formation and evolution was 
studied using numerical hydrodynamics simulations, while the formation and evolution of the
central star was calculated using a stellar evolution code. Three scenarios for mass accretion
from the disk onto the star were considered: hybrid accretion in which a fraction
of accreted energy absorbed by the protostar depends on
the accretion rate, hot accretion wherein a fraction of accreted energy is constant, 
and cold accretion wherein all accretion energy is radiated away.}
{Our conclusions on the nature of VeLLOs depend crucially on the character of protostellar 
accretion. In the hybrid accretion scenario, most
VeLLOs ($90.6\%$) are expected to be the first hydrostatic cores (FHSCs) and only 
a small fraction ($9.4\%$) are true protostars. In the hot accretion scenario, all VeLLOs are FHSCs
due to overly high photospheric luminosity of protostars. 
In the cold accretion scenario, on the contrary,
the majority of VeLLOs belong to the Class I phase of stellar evolution. The reason 
is that the stellar photospheric luminosity, which sets the floor for the total 
internal luminosity of a young star, is lower in cold accretion, thus enabling more VeLLOs
in the protostellar stage.  
VeLLOs are relatively rare objects occupying 7\%--11\% of the total duration of the embedded phase
and their masses do not exceed $0.3~M_\odot$.
When compared with published observations inferring a fraction 
of VeLLOs in the protostellar stage of $\sim 6.25\%$, we find that 
cold accretion provides a much better fit to observations than hybrid accretion
(5.7\% for cold accretion vs. 0.7\% for hybrid accretion). 
Both accretion scenarios predict more VeLLOs in the 
Class I phase than in the Class 0 phase, in contrast to observations.
Finally, when accretion variability with episodic bursts is artificially filtered out 
from our numerically derived accretion rates, the fraction of VeLLOs in the protostellar stage 
drops significantly, suggesting a causal link between the two phenomena. }
{}
 
\authorrunning{Vorobyov et al.}
\titlerunning{The nature of VeLLOs}
\maketitle

\section{Introduction}

Low-mass stars form when dense molecular cloud cores become unstable under their 
own self-gravity and collapse.  The process of low-mass star formation has 
received considerable attention in the last few decades, leading to the 
emergence of a general picture of the evolution from dense core to star 
\citep[e.g.,][]{shu1987:review,mckee2007:review}.  However, despite the 
existence of this general picture, a detailed physical understanding of 
the accretion process during the protostellar stage, the stage when the 
forming star is still embedded within and accreting from its parent 
dense core, remains elusive.  In the simplest model of star formation, 
the collapse of a singular isothermal sphere (sometimes called the 
standard model
of star formation), mass accretes from the core onto the 
protostar at a constant rate of $\sim$$2~\times~10^{-6}$~\msun~yr$^{-1}$ 
\citep{shu1977:sis,shu1987:review}.  However, most protostars have luminosities 
significantly less than the accretion luminosity expected from accretion at 
this rate, as first discussed by \citet{kenyon1990:lum}.  This luminosity problem, as it is called, has been emphasized by results 
from \textit{Spitzer Space Telescope} \citep{werner2004:spitzer} 
observations of nearby star-forming 
regions, which find a significant population of protostars with luminosities 
below theoretical predictions \citep[e.g.,][]{Dunham2008,Dunham2013,Dunham2014,
Dunham2015,Evans2009,enoch2009:protostars,kryukova2012:luminosities}.

There is growing evidence that the resolution of the luminosity problem lies 
in episodic rather than constant mass accretion, with prolonged periods of 
very low accretion punctuated by short bursts of rapid accretion.  
\citet{VB2005,VB2006,VB2010,VB2015} presented simulations showing that 
material piles up in a circumstellar disk until the disk becomes 
gravitationally unstable and dumps its mass in the form of gaseous clumps 
onto the protostar in short-lived 
accretion bursts. Indeed, the luminosities predicted by these simulations 
match the observed protostellar luminosity distribution and fully 
resolve the luminosity problem \citep{Dunham2010a,DV2012}.  While other 
accretion models are also capable of resolving the luminosity problem 
without invoking variability \citep{offner2011:luminosities}, significant 
indirect evidence exists in favor of episodic, variable mass accretion rates 
in the protostellar stage of star formation \citep{audard2014:ppvi,Dunham2014}, 
including infrared variability in young stars 
\citep[e.g.,][]{billot2012:variability,scholz2013:variability,rebull2014:ysovar,rebull2015:ysovar,gunther2014:ysovar}, outflow variability 
\citep[e.g.,][]{bachiller1991:outflows,arce2001:outflows,arce2013:outflows,Dunham2006,Dunham2010,Lee2010:outflows,Schwarz2012:outflows}, and chemical indications of extreme temperature variations in the histories of protostellar cores 
\citep[e.g.,][]{kim2012:chemistry,visser2012:chemistry,visser2015:chemistry,jorgensen2013:chemistry,vorobyov2013:chemistry}.

\subsection{Very low luminosity objects}

The very first dense molecular cloud core to be observed by \spitzer\ that 
was classified as starless (Lynds 1014), and was in fact revealed to harbor a very 
low luminosity protostar \citep{young2004:l1014}.  Several other, similar 
detections quickly followed 
\citep[e.g.,][]{Dunham2006,Dunham2010,bourke2006:l1521f,lee2009:l328}, 
leading to the definition of a new class of objects called Very Low Luminosity 
Objects \citep[VeLLOs;][]{DF2007:ppv}.  Defining the internal luminosity of 
a protostar, \lint, as the total luminosity arising from the star and disk 
(including both accretion and photospheric luminosity) and excluding any 
luminosity arising from external heating of the surrounding dense core by the 
interstellar radiation field, VeLLOs are defined as protostars with 
\lint~$<$~0.1~\lsun\ embedded in dense cores.  In a comprehensive search for 
VeLLOs in \spitzer\ observations of nearby star-forming regions, 
\citet{Dunham2008} found 15 total VeLLOs, 8 in isolated globules and 7 in 
large molecular cloud complexes.  In a related study, \citet{Evans2009} 
identified a total of 112 protostars in these same large molecular cloud 
complexes.  Thus VeLLOs represent 6.25\% (7/112) of the total population of 
protostars. We, however, acknowledge that this value is likely a 
lower limit to the true fraction. 

More recently, a number of embedded objects below the sensitivity of these 
large \spitzer\ surveys have been found, most through either interferometric 
detections of outflows driven by cores starless even via observations made by \spitzer, 
or far-infrared detections of warm dust toward such cores 
\citep{chen2010:fhsc,chen2012:fhsc,enoch2010:fhsc,dunham2011:fhsc,pineda2011:fhsc,pezzuto2012:fhsc,schnee2012:fhsc,murillo2013:fhsc}.  
With typical values of \lint~$\le$~0.01~\lsun, these objects have been 
suggested as candidates for the first hydrostatic core (FHSC), a short-lived 
stage (0.5--50~kyr according to \citet{Omukai2007,Tomida2010,Commercon2012}) 
intermediate between the starless and protostellar stages that forms once 
the central density increases to the point where the inner region becomes 
opaque to radiation and lasts until the central temperature reaches 2000~K 
and H$_2$ dissociates \citep{larson1969:fhsc}.  
Since most of these extreme objects have been discovered serendipitously, and 
complete surveys 
with the required sensitivity to find all of them are only 
just starting to become available \citep{dunham2016:chami}, and, further, since we do not yet have 
clear observational methods for distinguishing between first cores and very young protostars, 
the total number of these objects and their input to the VeLLO fractions (FHSCs vs. protostars) 
remains to be understood.




While VeLLOs thus represent less than approximately 10\% of the total population of 
protostars in nearby molecular clouds, their very low luminosities are an 
extreme example of the luminosity problem discussed above and are particularly 
challenging to explain.  Assuming accretion at the standard rate of 
$\sim$2$~\times~10^{-6}$~\msun~yr$^{-1}$ \citep{shu1987:review} onto a star on
the stellar/substellar boundary (0.08~\msun) with a typical protostellar 
radius of 3 \rsun, the expected accretion luminosity is 
$L=GM\dot{M}/R=1.6$~\lsun.  VeLLOs, with luminosities more than an order of 
magnitude below this, must feature some combination of very low masses and/or 
mass accretion rates.  While both have been indirectly inferred in detailed 
studies of individual VeLLOs 
\citep{Dunham2006,Dunham2010,lee2009:l328,lee2013:l328}, to date, no method has 
directly measured either their masses or mass accretion rates, since 
VeLLOs are too deeply embedded for optical/NIR spectroscopic measurements 
of accretion rates and spectral types.  Thus the physical nature of VeLLOs 
remains unknown.

\subsection{The scope of this work}

The purpose of this study is to develop a physical understanding of the nature 
of VeLLOs.  We use the hydrodynamical simulations of 
\citet{VB2005,VB2006,VB2010,VB2015}, which reproduce the full distribution of 
observed protostellar luminosities, including VeLLOs \citep{DV2012}, to 
investigate the characteristics of the protostars forming in these simulations 
with luminosities consistent with the definition of a VeLLO as given above.  
The organization of this work is as follows:  Section~\ref{model} presents 
the description of our numerical model and adopted initial conditions.
The main results are presented in Section~\ref{results}. The model
caveats are discussed in Section~\ref{caveats} and the conclusions are
given in Section~\ref{conclude}.

\section{Model description and initial conditions}

\label{model}
We start our numerical hydrodynamics simulations from the gravitational collapse 
of a starless cloud core, continue into the embedded phase of star formation, during which
a star, disk, and envelope are formed, and terminate our simulations after the parental 
core has dissipated (0.1-0.6~Myr, depending on the core mass). 
Such long integration times are made possible by the use
of the thin-disk approximation. This approximation is an excellent means to calculate
the evolution for many models and orbital periods and its justification is discussed
in \citet{VB2010}. The star, once formed, is placed in the coordinate center. 
The protostellar disk, once formed, occupies the inner part 
of the numerical polar grid (usually, several hundreds of AU), while the contracting 
envelope occupies the rest of the grid (which may extend to several thousands of AU).
The disk, during its early evolution, is not isolated, but is exposed to intense mass 
loading from the infalling envelope.  

To avoid overly small time steps, we introduce a `sink cell' 
near the coordinate origin with a radius of 5~AU 
and impose a free outflow boundary condition so that the matter is allowed to flow out of 
the computational domain into the sink cell, but is prevented from flowing 
in the opposite direction. 
During the early stages of the core collapse, we monitor the gas surface density in 
the sink cell and when its value exceeds a critical value for the transition from 
isothermal to adiabatic evolution ($\sim 10^{10}$~cm$^{-3}$), we assume that 
a first hydrostatic core (FHSC) is formed with a size 
equal to that of the sink cell. 
The second collapse ensues and a central protostar forms
when the mass of the FHSC exceeds 0.05~$M_\odot$,
as suggested by radiation transfer calculations of \citet{MI2000}. 
The mass contained in the FHSC (or in the sink cell) is then transferred to the point-sized 
protostar in the coordinate center over a short transition period of time (a few thousand years).
A small fraction of this mass (several per cent) remains in the sink cell
to guarantee a smooth transition of the gas surface density across the inner boundary. 
In the subsequent evolution, 90\% of the gas that crosses the inner boundary 
is assumed to land on the protostar. 
The other 10\% of the accreted gas is assumed to be carried away with protostellar jets. 


\subsection{Stellar evolution code}

The evolution of the accreting protostar is based on the Lyon stellar evolution code with input 
physics described in \cite{Chabrier97} including accretion processes as described in  
\citet{Baraffe09} and \citet{Baraffe12}. 
The accretion rates onto the protostar $\dot{M}$, calculated as the amount of mass passing through the
sink cell during one computational time step,  
are derived from the hydrodynamic calculations described below. 
As in \citet{Baraffe12}, we assume that the fraction $\alpha$ of the accretion energy  
$\epsilon G M_\ast{\dot M}/{R_\ast}$ is absorbed by the protostar, while the fraction 
$(1-\alpha$) is radiated away and contributes to the accretion luminosity of the star. As in \citet{Baraffe09}, we assume a value $\epsilon$=1/2 characteristic of accretion 
from a thin disk. The current mass of the star $M_\ast$ is calculated from the mass 
accretion rate $\dot{M}$, while the radius of the star $R_\ast$ is computed by the stellar 
evolution code.

Despite many efforts, the exact value of $\alpha$ in low-mass star formation is not 
known. In the present calculations, we consider three scenarios: a "cold" accretion model with
$\alpha=0$, meaning that  essentially all accretion energy is radiated away, 
a "hot" accretion model with $\alpha=0.5$, meaning that 50\% of the accretion energy
is always absorbed by the protostar, irrespective of the accretion rate, and a "hybrid" accretion
model with $\alpha=0$,  when accretion rates remain smaller than  a critical value 
$\dot{M}_{\rm cr}$ and  $\alpha \ne 0$  when $\dot{M} >  \dot{M}_{\rm cr}$.
More specifically, we adopt 
\begin{equation}
\label{alpha}
\alpha  = \left\{ \begin{array}{ll} 
   0,   &\,\,\, \mbox{if \, $\dot{M}<10^{-5} M_\odot$~yr$^{-1}$ }, \\ 
   0.2,  & \,\,\, \mbox{if \, $\dot{M}\ge 10^{-5} M_\odot$~yr$^{-1}$}. 
   \end{array} 
   \right. 
\end{equation}

The choice for the maximum value of $\alpha$ is based on our previous work showing
that stellar properties for hybrid accretion do not change significantly as long 
as $\alpha$ is greater than 
$0.1$--$0.2$ \citep{Baraffe09,Baraffe12}. The choice for the critical mass accretion rate of 
$\dot{M}_{\rm cr}=10^{-5}~M_\odot$~yr$^{-1}$ is based on our analysis of the pressure balance
at the stellar surface (see the Introduction and appendix B in \citet{Baraffe12}. 
For the initial seed mass of the protostar,  
corresponding to the second Larson core mass,  we adopt a value of 1.0$~M_{\rm
Jup}$ with an initial radius  $\sim 1.0~R_\odot$.

The stellar evolution  code is coupled with the main hydrodynamical code in real time 
(no postprocessing), meaning that we simultaneously model the evolution of 
the star and its circumstellar disk.
The input parameter to the stellar evolution code provided by disk modeling is 
the mass accretion rate onto the star $\dot{M}$.
The output of the stellar evolution code is the stellar radius $R_\ast$ and the photospheric 
luminosity $L_{\ast,\rm ph}$, which are employed by the disk hydrodynamics simulations 
to calculate 
the total stellar luminosity and the radiation flux reaching the disk surface (see below).
Due to heavy computational load,  the stellar evolution code is 
invoked to update the properties of the protostar 
only every 20~yr, while the hydrodynamical time step may be as small
as a few months. As a consequence, $\alpha$ has to be averaged over a  
period of 20~yr preceding the calculation of
the stellar properties. This means that
the actual value of $\alpha$ can be somewhat lower than the maximal value of 0.2.

\subsection{Numerical hydrodynamics code}
\label{numeric}
The main physical processes taken into account when computing the evolution of the 
disk and envelope include viscous and shock heating, irradiation by the forming star, 
background irradiation, radiative cooling from the disk surface and self-gravity. 
The corresponding equations of mass, momentum, and energy transport  are
\begin{equation}
\label{cont}
\frac{{\partial \Sigma }}{{\partial t}} =  - \nabla_p  \cdot 
\left( \Sigma \bl{v}_p \right),  
\end{equation}
\begin{eqnarray}
\label{mom}
\frac{\partial}{\partial t} \left( \Sigma \bl{v}_p \right) &+& \left[ \nabla \cdot \left( \Sigma \bl{v_p}
\otimes \bl{v}_p \right) \right]_p =   - \nabla_p {\cal P}  + \Sigma \, \bl{g}_p + \\ \nonumber
& + & (\nabla \cdot \mathbf{\Pi})_p, 
\label{energ}
\end{eqnarray}
\begin{equation}
\frac{\partial e}{\partial t} +\nabla_p \cdot \left( e \bl{v}_p \right) = -{\cal P} 
(\nabla_p \cdot \bl{v}_{p}) -\Lambda +\Gamma + 
\left(\nabla \bl{v}\right)_{pp^\prime}:\Pi_{pp^\prime}, 
\end{equation}
where subscripts $p$ and $p^\prime$ refer to the planar components $(r,\phi)$ 
in polar coordinates, $\Sigma$ is the mass surface density, $e$ is the internal energy per 
surface area, 
${\cal P}$ is the vertically integrated gas pressure calculated via the ideal equation of state 
as ${\cal P}=(\gamma-1) e$ with $\gamma=7/5$,
$\bl{v}_{p}=v_r \hat{\bl r}+ v_\phi \hat{\bl \phi}$ is the velocity in the
disk plane, and $\nabla_p=\hat{\bl r} \partial / \partial r + \hat{\bl \phi} r^{-1} 
\partial / \partial \phi $ is the gradient along the planar coordinates of the disk. 
The gravitational acceleration in the disk plane, $\bl{g}_{p}=g_r \hat{\bl r} +g_\phi \hat{\bl \phi}$, takes into account self-gravity of the disk, found by solving for the Poisson integral 
\citep[see details in][]{VB2010}, and the gravity of the FHSC and central protostar when formed. 
Turbulent viscosity due to sources other than gravity 
is taken into account via the viscous stress tensor 
$\mathbf{\Pi}$, the expression for which is provided in \citet{VB2010}.
We parameterize the magnitude of kinematic viscosity $\nu$ using the alpha prescription 
with a spatially and temporally uniform $\alpha_{\rm visk}=5\times 10^{-3}$.
The thin-disk model is complemented by a calculation of the vertical 
scale height $h$ in both the disk and envelope using an assumption of local 
hydrostatic equilibrium \citep{VB2009}. 
The resulting model has a flared structure, which guaranties that 
both the disk and envelope receive a fraction of the irradiation energy 
from the central protostar.

The radiative cooling $\Lambda$ in equation~(\ref{energ}) is determined using the diffusion
approximation of the vertical radiation transport in a one-zone model of the vertical disk 
structure \citep{Hubeny1990,JG2003}
\begin{equation}
\Lambda={\cal F}_{\rm c}\sigma\, T_{\rm mp}^4 \frac{\tau}{1+\tau^2},
\end{equation}
where $\sigma$ is the Stefan-Boltzmann constant, $T_{\rm mp}={\cal P} \mu / R \Sigma$ is 
the midplane temperature of gas, $\mu=2.33$ is the mean molecular weight, 
$R$ is the universal 
gas constant,  $\tau$ is the optical depth to the disk midplane, and ${\cal F}_{\rm c}=2+20\tan^{-1}(\tau)/(3\pi)$ is a function that 
secures a correct transition between the optically thick and optically thin regimes.  
We use frequency-integrated opacities of \citet{BL94} to calculate the optical depth.
The heating function is expressed by analogy to the cooling function as 
\begin{equation}
\Gamma={\cal F}_{\rm c}\sigma\, T_{\rm irr}^4 \frac{\tau}{1+\tau^2},
\end{equation}
where $T_{\rm irr}$ is the irradiation temperature at the disk surface 
determined by the stellar and background black-body irradiation as
\begin{equation}
T_{\rm irr}^4=T_{\rm bg}^4+\frac{F_{\rm irr}(r)}{\sigma},
\label{fluxCS}
\end{equation}
where $T_{\rm bg}$ is the uniform background temperature (in our model set to the 
initial temperature of the natal cloud core)
and $F_{\rm irr}(r)$ is the radiation flux (energy per unit time per unit surface area) 
absorbed by the disk surface at radial distance 
$r$ from the central object. The latter quantity is calculated as 
\begin{equation}
F_{\rm irr}(r)= \frac{L_{\rm obj}}{4\pi r^2} \cos{\gamma_{\rm irr}},
\label{fluxF}
\end{equation}
where $\gamma_{\rm irr}$ is the incidence angle of 
radiation arriving at the disk surface at radial distance $r$. The incidence angle is calculated
using the disk surface curvature inferred from the radial profile of the  
disk vertical scale height \citep{VB2010}.

The luminosity of the central object $L_{\rm obj}$ is the sum of the accretion luminosity
$L_{\rm accr}=(1-\alpha)\epsilon G M_{\rm obj} \dot{M}/R_{\rm obj}$ arising from the 
gravitational energy of accreted gas and
the photospheric luminosity $L_{\rm \ast,ph}$ due to gravitational contraction and
deuterium burning. We note that while the accretion luminosity is calculated for both
the FHSC and protostar, the photospheric luminosity is calculated only for the protostar (using the
stellar evolution code), but not for the FHSC (our simplistic model does not allow us to do this).
The mass of the central object $M_{\rm obj}$ and accretion rate onto the central object $\dot{M}$
are determined self-consistently during numerical simulations using the amount of gas passing through
the sink cell. We note that the radius of the central object is set to the radius of the sink cell in
the FHSC phase and is self-consistently determined by the stellar evolution code in the protostellar
phase. 
The numerical resolution is $512\times 512$ grid points and the numerical procedure to solve
hydrodynamics equations~(\ref{cont})-(\ref{energ})  is described in detail in \citet{VB2010}. 

\subsection{Initial setup}
Our initial cloud cores are described by the gas surface density $\Sigma$  
and angular velocity $\Omega$ profiles of the following form:  
\begin{equation}
\Sigma={r_0 \Sigma_0 \over \sqrt{r^2+r_0^2}}\:,
\label{dens}
\end{equation}
\begin{equation}
\Omega=2\Omega_0 \left( {r_0\over r}\right)^2 \left[\sqrt{1+\left({r\over r_0}\right)^2
} -1\right].
\label{omega}
\end{equation}
Here, $\Omega_0$ and $\Sigma_0$ are the angular velocity and gas surface
density at the center of the core and $r_0 =\sqrt{A} c_{\rm s}^2/\pi G \Sigma_0 $
is the radius of the central plateau, where $c_{\rm s}$ is the initial sound speed in the core. 
The gas surface density distribution described by equation~(\ref{dens}) 
is a vertically integrated form (to within a factor of order unity) of 
the Bonnor-Ebert sphere with a positive density-perturbation amplitude A \citep{Dapp2009}.
The value of $A$ is set to 1.2 and the initial gas temperature is set to 10~K.
We also note that our initial conditions are typical of pre-stellar cores formed 
as a result of the slow expulsion 
of magnetic field due to ambipolar diffusion, with the angular
momentum remaining constant during axially-symmetric core compression \citep{Basu97}

Our models are divided into four model sets, each characterized
by a distinct ratio of the rotational to gravitational energy $\beta$. The adopted values of 
$\beta$  lie within the limits inferred by \citet{Caselli2002} for dense molecular cloud cores. 
Each model is characterized by a distinct ratio $r_{\rm out}/r_0=6$ 
in order to generate gravitationally unstable truncated cores of similar form, where $r_{\rm out}$ is
the outer radius of the cloud core. The actual procedure for generating the parameters of 
a specific core with a given value of $\beta$ is as follows. First, we choose
 $r_{\rm out}$ and find $r_0$ using the adopted ratio between these two quantities.
Then, we find the central surface density $\Sigma_0$ from the relation 
$r_0=\sqrt{A}c_{\rm s}^2/(\pi G \Sigma_0)$ and determine the resulting cloud core mass
$M_{\rm c}$ from Equation~(\ref{dens}).

\begin{figure*}
\begin{centering}
\includegraphics[scale=0.8]{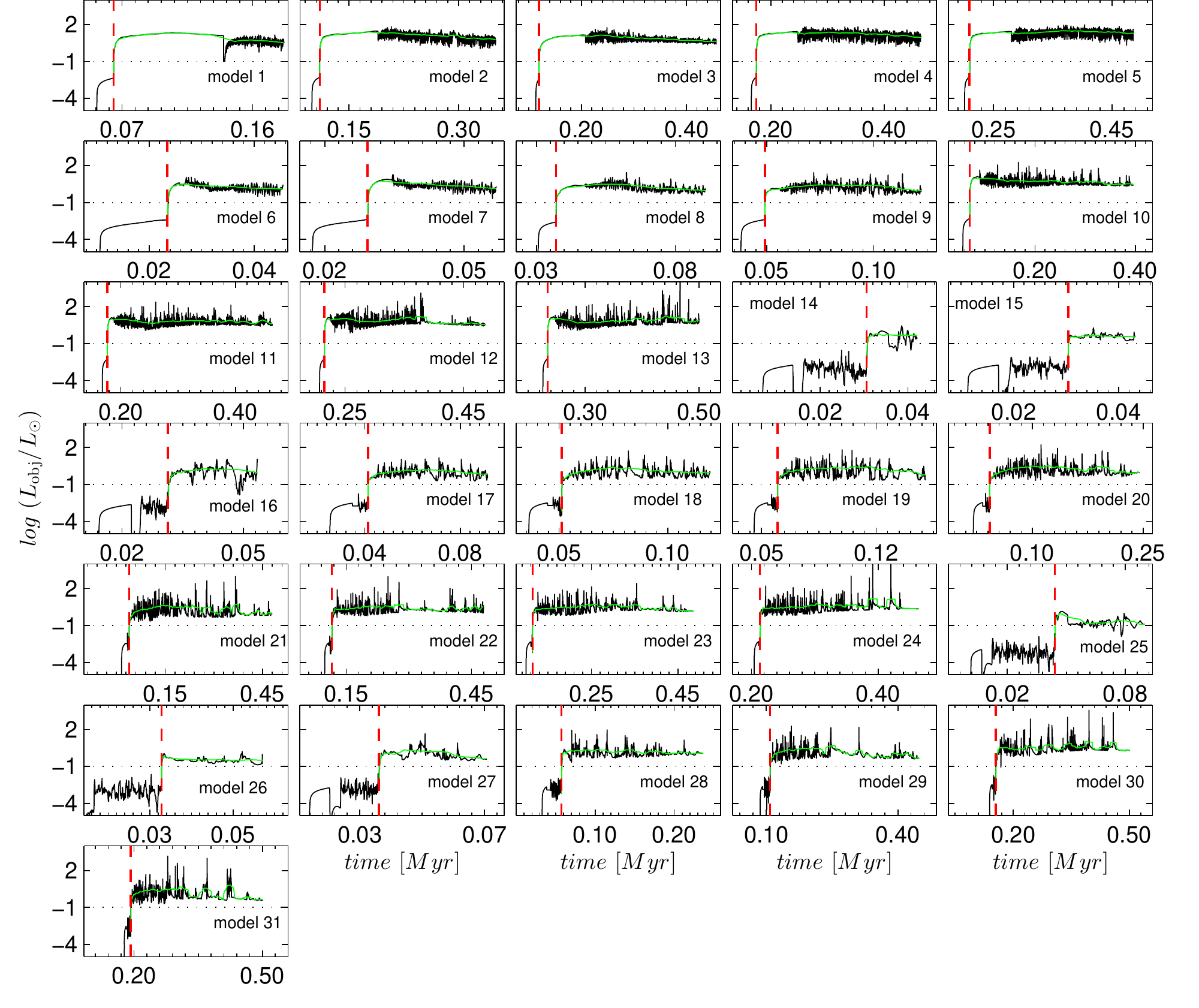}
\par\end{centering}
\protect\caption{\label{fig1} Total luminosity of the central object vs. time elapsed since the 
onset of core collapse in 31 models 
with the hybrid accretion scenario (black solid lines). The horizontal
dotted lines indicate the upper luminosity for VeLLOs.
The vertical dashed lines mark the instance of protostar formation.
The green solid lines present the total stellar luminosity averaged over a time period of 20~kyr
for models 21--24 and 29--31 and over a time period of 5~kyr for the remaining models.}
\end{figure*}

\section{Model results}
\label{results}
We have computed the evolution of our models to the end of the embedded phase,
when most of the collapsing core has dissipated via accretion onto
the forming disk plus star system. The numerical simulations are  terminated when 
the mass in the infalling envelope becomes smaller than 10\% 
of that of the initial core mass $M_{\rm c}$. We use the obtained total (accretion plus
photospheric) luminosities of the central objects to infer the nature of VeLLOs.

\subsection{Hybrid accretion}
\label{hybridAR}

\begin{table}
\protect\caption{\label{table1}Parameters for the hybrid accretion models}
\begin{centering}
\begin{tabular}{ccccc}
\hline 
Model & $M_{\mathrm{c}}$ & $\beta$  & $r_{\mathrm{out}}$ & $M_{*,\,\mathrm{fin}}$\tabularnewline
 & {[}$M_{\odot}${]} & {[}\%{]} & {[}pc{]} & {[}$M_{\odot}${]}\tabularnewline
\hline 
1 & 0.46 & 0.14 & 0.03 & 0.34\tabularnewline
2 & 0.92 & 0.14 & 0.06 & 0.57\tabularnewline
3 & 1.15 & 0.14 & 0.075 & 0.65\tabularnewline
4 & 1.38 & 0.14 & 0.09 & 0.78\tabularnewline
5 & 1.69 & 0.14 & 0.11 & 0.82\tabularnewline
6 & 0.11 & 0.57 & 0.007 & 0.07\tabularnewline
7 & 0.14 & 0.57 & 0.009 & 0.09\tabularnewline
8 & 0.23 & 0.57 & 0.015 & 0.12\tabularnewline
9 & 0.32 & 0.57 & 0.02 & 0.17\tabularnewline
10 & 0.94 & 0.57 & 0.06 & 0.50\tabularnewline
11 & 1.38 & 0.57 & 0.09 & 0.67\tabularnewline
12 & 1.71 & 0.57 & 0.11 & 0.61\tabularnewline
13 & 1.92 & 0.57 & 0.125 & 0.72\tabularnewline
14 & 0.09 & 1.74 & 0.006 & 0.05\tabularnewline
15 & 0.11 & 1.74 & 0.007 & 0.06\tabularnewline
16 & 0.12 & 1.74 & 0.008 & 0.06\tabularnewline
17 & 0.23 & 1.74 & 0.015 & 0.12\tabularnewline
18 & 0.31 & 1.74 & 0.02 & 0.15\tabularnewline
19 & 0.35 & 1.74 & 0.023 & 0.19\tabularnewline
20 & 0.46 & 1.74 & 0.03 & 0.25\tabularnewline
21 & 0.69 & 1.74 & 0.045 & 0.39\tabularnewline
22 & 0.94 & 1.74 & 0.06 & 0.43\tabularnewline
23 & 1.21 & 1.74 & 0.08 & 0.45\tabularnewline
24 & 1.54 & 1.74 & 0.1 & 0.58\tabularnewline
25 & 0.09 & 2.87 & 0.006 & 0.052\tabularnewline
26 & 0.12 & 2.87 & 0.008 & 0.06\tabularnewline
27 & 0.15 & 2.87 & 0.01 & 0.07\tabularnewline
28 & 0.46 & 2.87 & 0.03 & 0.26\tabularnewline
29 & 0.69  & 2.87 & 0.045 & 0.32\tabularnewline
30 & 1.23 & 2.87 & 0.08 & 0.57\tabularnewline
31 & 1.38 & 2.87 & 0.09 & 0.32\tabularnewline
\hline 
\end{tabular}
\par\end{centering}
\end{table}

In this section, we present our model results for the hybrid accretion scenario, wherein 
the fraction of accretion energy $\alpha$ absorbed by the protostar depends on the protostellar
accretion rate and is determined using equation~(\ref{alpha}). 
We use 31 models, the parameters of which are presented in Table~\ref{table1}. 
Columns 2--5 give the initial mass of the core, 
initial ratio of rotational to gravitational energy of
the core, initial outer radius of the core, and mass of the central object at the end of the embedded
phase, respectively.

We start by analyzing the total luminosity of the central object $L_{\rm obj}$ in
our models. The black solid lines in Figure \ref{fig1} present $L_{\rm obj}$ vs. time elapsed
since the beginning of the gravitational collapse of pre-stellar cores. 
The horizontal dotted lines indicate, according to \citet{Dunham2006}, a fiducial maximum 
value for the total luminosity of VeLLOs, $L^{\rm vello}_{\rm max}=0.1\, L_{\odot}$. 
The vertical dashed lines show the time instance when the FHSC collapses to form a
protostellar/proto-brown dwarf seed. 

In the FHSC phase, the total luminosity in all models is much smaller than
$L^{\rm vello}_{\rm max}$.  This is because the radius of the FHSC 
is at least three orders of magnitude larger than that of the protostar. 
Although the accretion rate in the FHSC phase (a~few~$\times 10^{-6}~M_\odot$~yr$^{-1}$)
is on average comparable to or even higher than that in the later protostellar phase
($10^{-7}-10^{-6}~M_\odot$~yr$^{-1}$), the resulting accretion luminosity in the FHSC phase is 
much smaller than in the protostellar phase.

When the protostar forms, $L_{\rm obj}$ rises by several orders of magnitude, initially exceeding 
the VeLLO's upper limit of $0.1~L_\odot$ in all models. 
In the subsequent evolution, the total stellar luminosity shows time variations with 
different amplitude. 
More specifically, $L_{\rm obj}$ in the low-$M_{\rm c}$ and low-$\beta$ models
(e.g., models 1--7) is characterized by an order-of magnitude
flickering. On the other hand, models
with higher $M_{\rm c}$ and $\beta$ (e.g., models 9--13,
17--24, 27--31) demonstrate large-amplitude variations
in $L_{\rm obj}$ with strong luminosity outbursts. 
This difference in the time behavior of $L_{\rm obj}$ stems
from the different properties of protostellar disks formed from the
gravitational collapse of pre-stellar cores \citep[e.g.,][]{VB2010}. 
The low-$M_{\rm c}$ and low-$\beta$ models produce disks of
low mass and size, which are weakly gravitationally unstable and show
no sign of fragmentation, while higher $M_{\rm c}$ and $\beta$ models
form disks that are sufficiently massive and extended to develop strong
gravitational instability and fragmentation. Most of the fragments
migrate onto the star owing to the loss of angular momentum
via gravitational interaction with spiral arms or other fragments
in the disk, producing strong accretion and luminosity bursts similar in magnitude
to FU-Orionis-type eruptions \citep{VB2006,VB2010,Machida2011,VB2015}.

\begin{figure}
  \resizebox{\hsize}{!}{\includegraphics{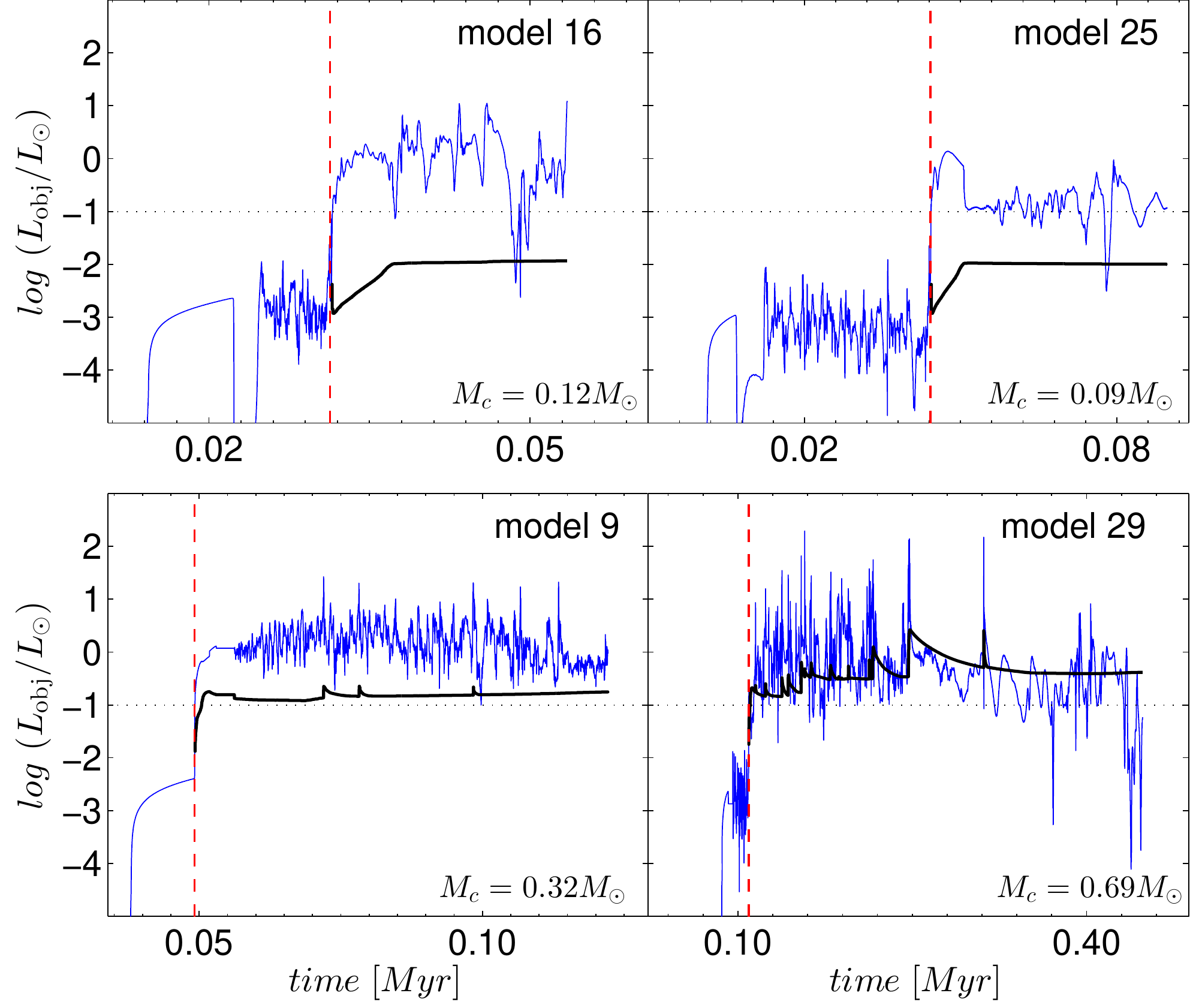}}
  \caption{Accretion and photospheric luminosities (thin blue and thick black lines, respectively)
  vs. time elapsed since the onset of core collapse in four prototype models. The vertical 
  red dashed line indicates the beginning of the protostellar phase (i.e., the onset of the
   collapse of a FHSC to a protostellar seed). The horizontal dotted lines indicate the upper luminosity for VeLLOs.}
  \label{fig2}
\end{figure}

Notwithstanding high variability, the total stellar luminosity in the protostellar phase 
in most models is greater than the VeLLO's upper limit of 0.1~$L_\odot$. Although the accretion luminosity
may drop below 0.1~$L_\odot$ following large-amplitude variations in the protostellar accretion rate,
the photospheric luminosity is greater than 0.1~$L_\odot$ in most cases. 
We illustrate this phenomenon in 
Figure~\ref{fig2} for several prototype models.  In particular, the thin blue lines and thick black lines show the accretion and photospheric luminosities, respectively, while the vertical red dashed line separates 
the FHSC and protostellar phases. The horizontal dotted lines mark the 
maximum luminosity of VeLLOs, $L^{\rm vello}_{\rm max}=0.1~L_\odot$.
In a few models, such as models 16 and 25, characterized by low-$M_{\rm c}$ and low-$\beta$
cores, the protostellar accretion rates never exceed $\dot{M}_{\rm cr}=10^{-5}~M_\odot$~yr$^{-1}$, implying
essentially cold accretion. As a result, the protostar remains compact and its photospheric luminosity
is smaller than $L^{\rm vello}_{\rm max}$ \citep{Baraffe09,Baraffe12}. 
Large variations in the accretion rate and the corresponding luminosity then lead to short episodes
when the total protostellar accretion falls below the VeLLO's upper limit.

On the other hand, the photospheric luminosity in models 9 and 29 is always greater than 
$L^{\rm vello}_{\rm max}$.
In these (and many other) models, the protostellar accretion rates often exceed the critical value of
$10^{-5}~M_\odot$~yr$^{-1}$, resulting in absorption of part of the accretion energy by the protostar
and causing the star to bloat \citep{Baraffe12}.
These models are characterized by photospheric luminosity that is greater than the upper VeLLO's
limit of $L^{\rm vello}_{\rm max}=0.1~L_\odot$. This means that the total luminosity in these models is always greater than $L^{\rm vello}_{\rm max}$, no matter how low the accretion luminosity 
may actually drop.

\subsubsection{Statistical analysis}
\label{statHA}

To further clarify the nature of VeLLOs, 
we calculate the time spent by our models in the VeLLO state. 
For this purpose, we employ the method described by \citet{DV2012}
and determine the fraction of total time $f_{\rm bin}$  that all our models spend
in various bins in the $L_{\rm obj}$--$t_{\rm evol}$ diagram, where $t_{\rm evol}$
is the evolution time elapsed since the beginning of core collapse.
To calculate this fraction, we first divide the $L_{\rm obj}$--$t_{\rm evol}$
space into 20 bins in both dimensions (in the  log space) and then calculate
the fraction of total time spent by the models in each bin:
\begin{equation}
f\mathrm{_{bin}}=\frac{ \sum \limits^{31}_{i=1} t_{i}^{\mathrm{bin}} \omega_{i}}
{\sum \limits^{31}_{i=1} t_{i}^{\mathrm{total}}\omega_{i}},
\label{eq:1}
\end{equation}
where $t_{i}^{\mathrm{bin}}$ is the time spent by the \textit{i}th
model in the specified $L_{\rm obj}$--$t_{\rm evol}$ bin, 
$t_{i}^{\mathrm{total}}$ is the total duration of the \textit{i}th model, 
$\omega_{i}$ are the weight coefficients given to the \textit{i}th model
according to the adopted initial mass function (IMF), and the summation 
is performed over all 31 models.

\begin{figure}
  \resizebox{\hsize}{!}{\includegraphics{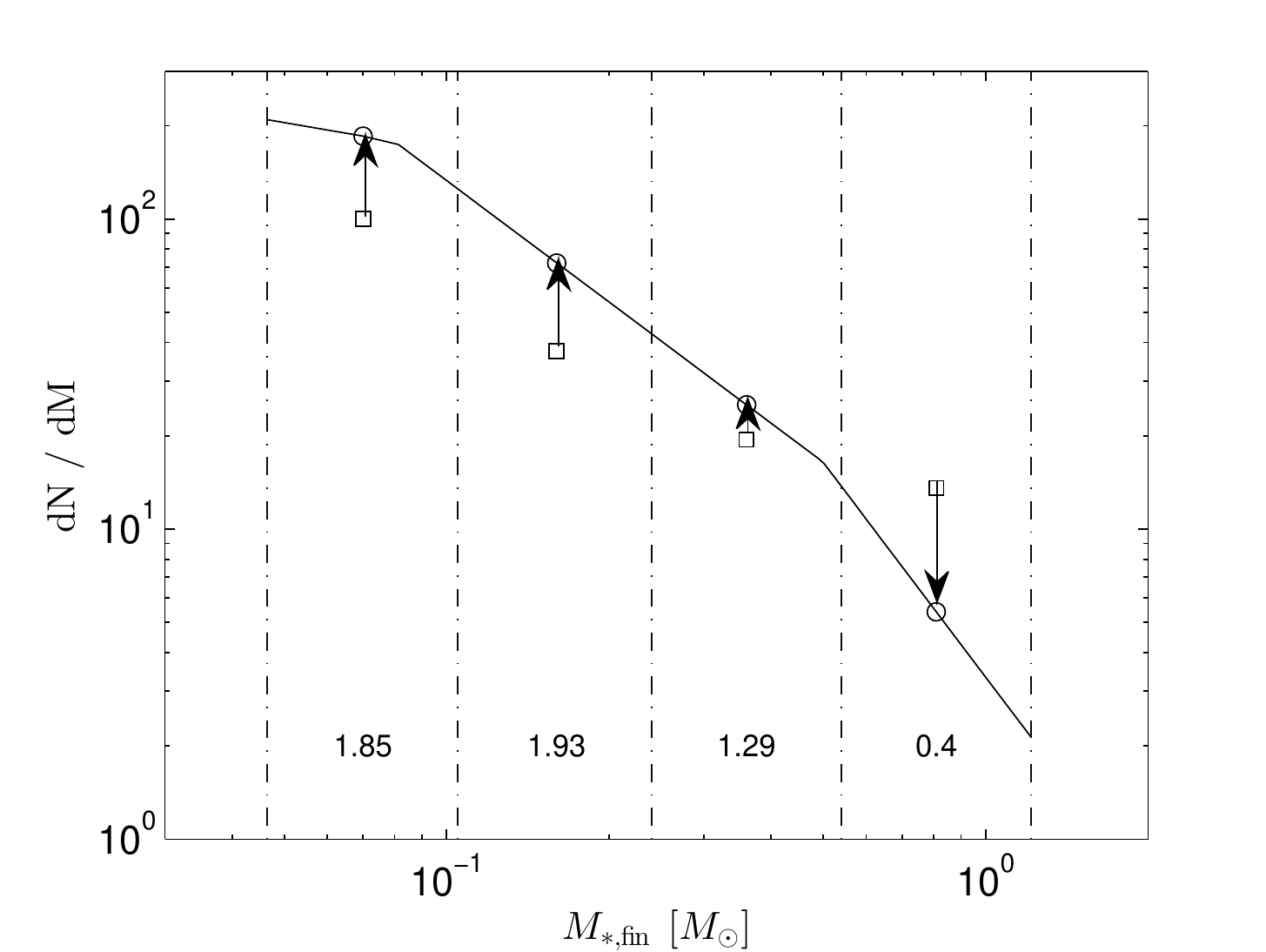}}
  \caption{Number of models in four mass bins ($dN/dM_{\ast, \rm fin}$) and the calculation of the weight
coefficients for the model IMF.  The
vertical dash-dotted lines outline the four bins, in which the final stellar masses 
($M_{\ast, \rm fin}$) are divided.
The solid line presents the Kroupa IMF normalized to the total number of our models. 
The open squares
and circles are the values of the model IMF before and after weighing, respectively,
and the numbers provide the weight coefficients $\omega_{i}$.}
  \label{fig3}
\end{figure}

To calculate  $\omega_{i}$, we split
the range of final stellar masses $M_{\rm \ast, fin}$ into four bins (in the log space)
and calculate the number of models per each mass bin, $dN/dM_{\ast, \rm fin}$. 
 The resulting model IMF is shown in Figure 
\ref{fig3} by the open squares. The mass bins are outlined in the figure by the vertical
dash-dotted lines.  The solid curve shows the Kroupa IMF 
\citep{Kroupa2001}, which is normalized to the total number of our models. 
We have also tried the Chabrier IMF \citep{Chabrier2005}, but found that the
adopted form of the IMF does not qualitatively affect our conclusions.
Evidently, the model IMF and the Kroupa IMF do not match, meaning 
that we have taken too few low-mass model cores and too
many high-mass ones in Table~\ref{table1}. This mismatch is caused
by the fact that several models have failed due to numerical reasons. The 
stellar evolution code sometimes diverges, which causes the whole 
simulation to be terminated.
We can, however, recover the Kroupa IMF by calculating 
the weight coefficients $\omega_{i}$ so as to fit our model
IMF to that of Kroupa. The arrows and open circles show the result
of the fitting and the numbers provide the derived weights for every
stellar mass bin. We note that we used the initial mass function of stars
for calculating the weights, rather than the initial mass function of 
dense molecular cores \citep[e.g.,][]{Alves2007}. This is because the latter sample
may include not only the pre-stellar, gravitationally contracting cores, 
as in our initial numerical setup, but also starless gravitationally stable cores.
In addition, some cores in the Alves et al. sample may undergo further fragmentation
when contracting to form binary/multiple stars or be partly photoevaporated/dispersed by 
stellar feedback. All these processes are not taken into account in our numerical
simulations of isolated cores.

\begin{figure}
  \resizebox{\hsize}{!}{\includegraphics{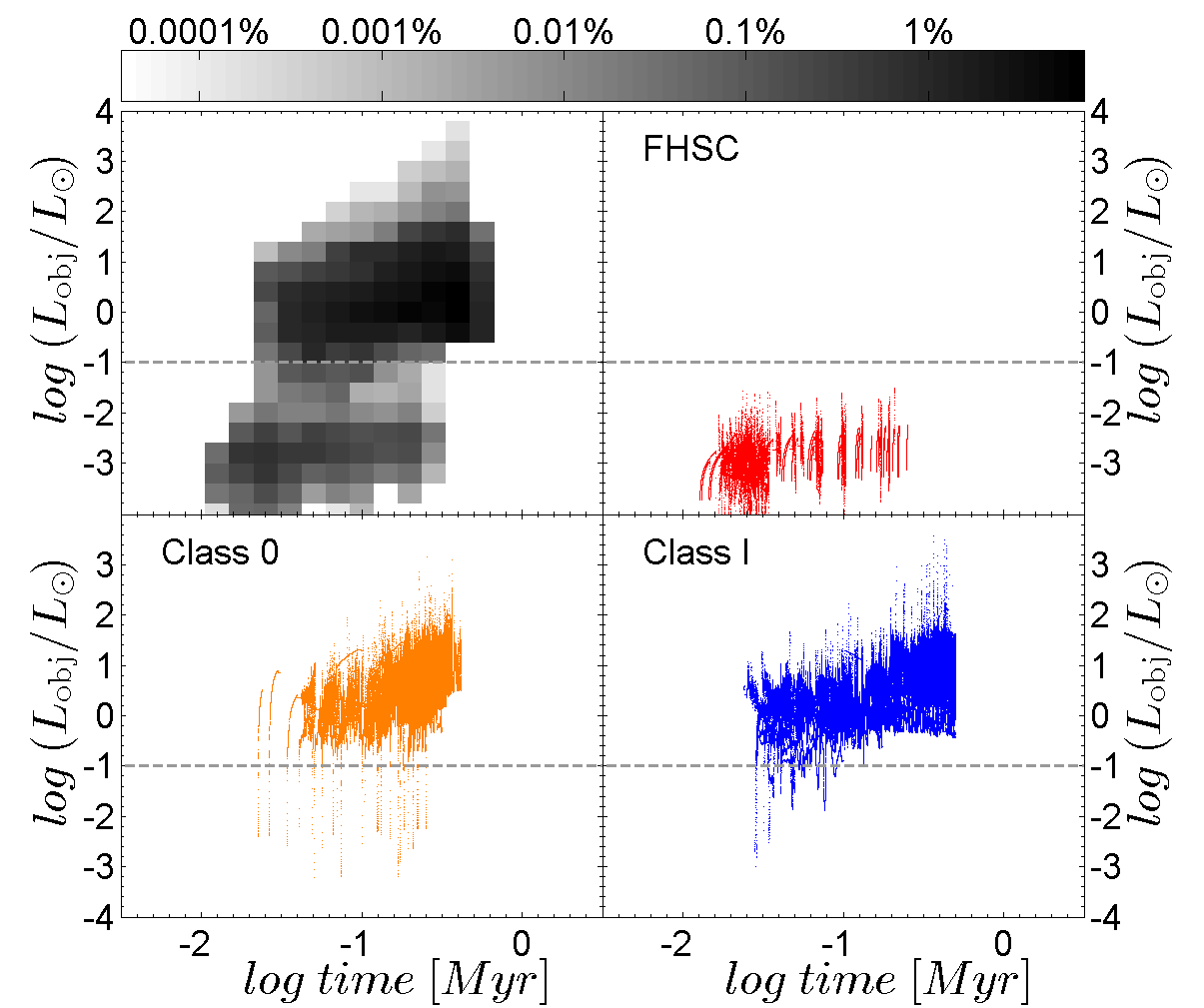}}
  \caption{Total luminosity ($L_{\rm obj}$)  vs. evolutionary time ($t_{\rm evol}$) diagram. 
  The time is counted from the onset
  of core collapse. The horizontal dashed lines mark the maximum VeLLO luminosity of 
  $0.1~L_\odot$. The gray-scale mosaic presents the fraction of total time (in per cent) 
  that 
  all models spend in various bins in the $L_{\rm obj}$--$t_{\rm evol}$ diagram. The corresponding
  scale bar shows this fraction in per cent. The red, orange, and blue dots 
  present the model data for 
  the FHSC, Class 0 and Class I phases, respectively.}
  \label{fig4}
\end{figure}

The fraction of total time $f_{\rm bin}$ is plotted in the upper-left panel of Figure~\ref{fig4}
with the gray-scale mosaic. Evidently, our models spend most of the evolution time in a
state with $L_{\rm obj}> L_{\rm max}^{\rm vellos}$, but VeLLOs with 
$L_{\rm obj} < L_{\rm max}^{\rm vellos}$  are also present. The gray-scale mosaic, however, 
gives little information regarding the evolutionary status of VeLLOs, that is, whether VeLLOs
are FHSCs or protostars.

To determine the evolutionary phase to which VeLLOs belong, we split our model data 
into three phases: the FHSC phase, the Class 0 phase, and
the Class I phase. The first phase starts from
the time instance of the FHSC formation and ends when the FHSC collapses 
to form the protostar. The second phase starts
from the moment of protostar formation and  ends when 50\% of the initial core mass 
is left in the envelope (the rest is accreted onto the star plus disk system). 
The third phase starts right after the Class 0 phase and ends 
when less than 10\% of the initial core mass is left in the envelope.
In all models, we assume that the Class I phase cannot last longer than 0.5~Myr, 
which is consistent with the lifetime of the 
embedded phase estimated from observations, 0.4--0.78~Myr \citep{Dunham2015}.
Our adopted classification scheme is therefore based on physical
properties of a young stellar object, such as envelope, disk, and stellar masses
\citep{Robitaille2006,Dunham2010}, and is more suitable for numerical simulations 
than observational signatures, such as submillimeter luminosity or
effective temperature \citep{Andre1993, Chen1995}. We note that we have adopted the
"Class" rather than "Stage" terminology in this paper, but the above clarification 
helps to avoid any confusions.

\begin{table}
\renewcommand{\arraystretch}{2.0}
\begin{centering}
{\small{}}%
\begin{tabular}{|c|c|c|c|}
\multicolumn{1}{c}{} & \multicolumn{3}{c}{}\tabularnewline
\hline 
Phase & $\frac{t_{\rm VeLLOs}^{\rm phase}}{t_{\rm VeLLOs}^{\rm tot}}$ & 
$\frac{t_{\rm VeLLOs}^{\rm phase}}{t_{\rm phase}}$ & $\frac{t_{\rm VeLLOs}^{\rm phase}}{t_{\rm
tot}}$\tabularnewline
\hline 
{\small{}FHSC} & {\small{}90.63} & {\small{}100.0} & {\small{}6.62}\tabularnewline
\hline 
{\small{}Class 0} & {\small{}0.69} & {\small{}0.21} & {\small{}0.05}\tabularnewline
\hline 
{\small{}Class I} & {\small{}8.68} & {\small{}0.89} & {\small{}0.63}\tabularnewline
\hline 
\end{tabular}
\par\end{centering}{\small \par}
\protect\caption{\label{table2}Fraction of time (in per cent) spent by hybrid accretion models in each evolution phase.  }
\end{table}

The red, orange, and blue dots in Figure~\ref{fig4} present the model data for 
the FHSC, Class 0 and Class I phases, respectively. The corresponding total
luminosities are calculated every 5~yr.
The dashed horizontal line shows the VeLLO's maximum luminosity, 
$L^{\rm vello}_{\rm max}=0.1~L_\odot$.
Evidently, the FHSC phase is characterized by total luminosity that is
much smaller than $L^{\rm vello}_{\rm max}$, which is not surprising considering 
the large radii and low accretion luminosities  of the FHSCs.  
Indeed, for a radius of 5~AU, maximum mass of 
0.05~$M_\odot$ and maximum accretion rate of 
$10^{-5}~M_\odot$~yr$^{-1}$, we obtain $1.2\times 10^{-2}~L_\odot$ for 
the maximum accretion luminosity of FHSCs, 
which agrees with the values in Figure~\ref{fig2} (but see discussion
in Section~\ref{caveats}).
The Class 0 and I phases, on the contrary, are mostly characterized 
by $L_{\rm obj}>L^{\rm vello}_{\rm max}$. The total luminosity drops 
to the VeLLO domain only episodically.
 
Using the constructed mosaic, we have further calculated the fractions of time
($t_{\rm VeLLOs}^{\rm phase}$) 
spent by our models in the VeLLO state with $L_{\rm obj}<0.1L_\odot$ 
during each of the three phases (FHSC, Class 0, and Class I) with respect to:
{a)} the total time spent with $L<0.1L_{\odot}$ during all three phases 
($t_{\rm VeLLOs}^{\rm tot}$), {b)} the total duration for each evolution phase 
($t_{\rm phase}$), and {c)} the total duration for all phases
($t_{\rm tot}$). The resulting values are shown in Table~\ref{table2}.

\begin{figure}
  \resizebox{\hsize}{!}{\includegraphics{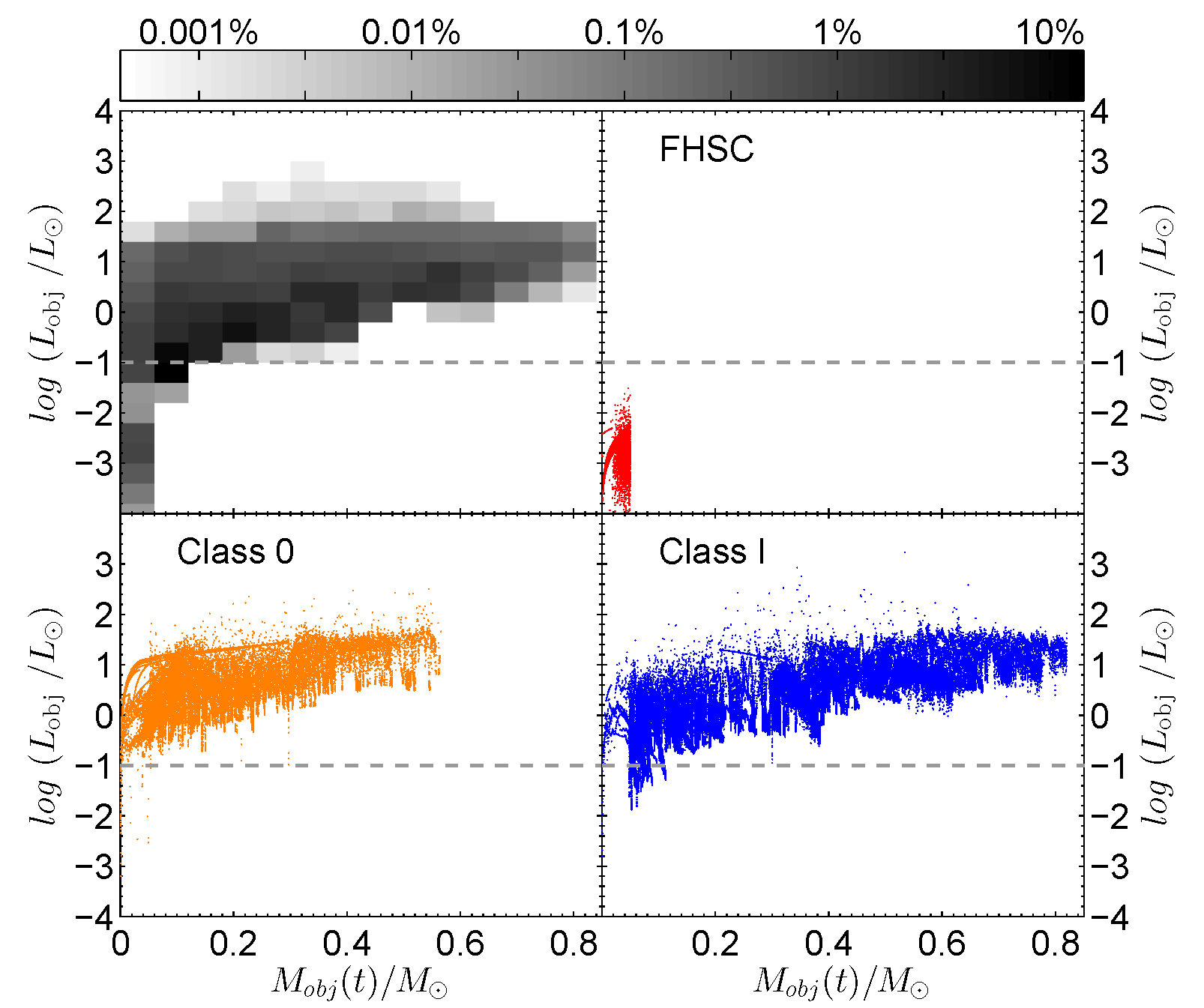}}
  \caption{Total luminosity ($L_{\rm obj}$)  vs. central object mass ($M_{\rm obj}$) diagram. 
  The horizontal dashed line marks the maximum VeLLO luminosity of 
  $0.1~L_\odot$. The gray-scale mosaic presents the fraction of total time (in per cent) 
  that 
  all models spend in various bins in the $L_{\rm obj}$--$M_{\rm obj}$ diagram. The corresponding
  scale bar shows this fraction in per cent. The red, orange, and blue dots 
  in Figure~\ref{fig2} present the model data for 
  the FHSC, Class 0 and Class I phases, respectively.}
  \label{fig5}
\end{figure}

To convert our numerically predicted time fractions to number fractions,
we make a fundamental assumption of a constant star formation rate. A similar
assumption is used, for example, when deriving the duration of 
different evolutionary phases from the observationally inferred 
number fractions of protostars  \citep[e.g.,][]{Dunham2015}.
Let us focus on the first numerical column showing the time 
spent in the VeLLO state for each individual phase 
normalized to the total time spent in the VeLLOs state  ($t_{\rm VeLLOs}^{\rm tot}$) for all 
three phases. These data can be used to derive the number of VeLLOs  in a given
phase if the total number of VeLLOs, $N^{\rm tot}_{\rm VeLLOs}$, in a given star-forming region is known:
\begin{equation}
N_{\rm VeLLOs} = 0.01 \times { t^{\rm phase}_{\rm VeLLOs} \over t^{\rm tot}_{\rm VeLLOs} } \times 
N^{\rm tot}_{\rm VeLLOs}.\end{equation}

Evidently, most VeLLOs are expected to be found in 
the FHSC phase ($90.63\%$) and only a small fraction can be
in the Class 0 ($0.69\%$) and Class~I phases ($8.68\%$).
This suggests that most VeLLOs are in fact the FHSCs,
and only a minor fraction are true protostars. 
We note that $N^{\rm tot}_{\rm VeLLOs}$ is a poorly constrained number
from observations. Nevertheless, the second numerical column 
provides an important theoretical insight into the nature of VeLLOs.


The second numerical column provides the time spent by our models in the VeLLO state  
in each individual evolution phase normalized to the duration of the phase. 
These data can be directly used to calculate the number of
VeLLOs as follows:
\begin{equation}
\label{Nvello}
N_{\rm VeLLOs} = 0.01 \times { t^{\rm phase}_{\rm VeLLOs} \over t_{\rm phase} } \times N_\ast,
\end{equation}
where $N_\ast$ is the number of  protostars in a specific phase and specific star-forming 
region. We note that Equation~(\ref{Nvello}) is only applicable to the protostellar phase.
This quantity is much better constrained from observations than $N^{\rm tot}_{\rm VeLLOs}$. 
From the third numerical column
it is evident that the FHSCs spend all their time in the VeLLO state,
as was already noted in Figure~\ref{fig4}. On the other hand,
Class 0 and Class I protostars spend only a very small
fraction of the lifetime of the corresponding phase in the VeLLO state, meaning
that it would be unlikely (from a statistical point of view)  
to detect Class 0 and Class I protostars in the VeLLO state
and their number count should be quite low, less than one per 
hundred protostars. This, however, contradicts observations.
As mentioned in the Introduction,  VeLLOs represent
6.25\% (7/112) of the total population of protostars (and perhaps more,
as discussed in Section~\ref{conclude}).

The third numerical column indicates that VeLLOs are relatively rare objects 
when compared to the total duration of the embedded
phase, including the FHSC phase and both Class 0 and I phases.
Although the total fraction of VeLLOs in all three phases (7.3\%) is similar to the observationally
inferred 6.25\%, this comparison is misleading because the {\it Spitzer} observations
in the 3.6--8 $\mu$m cannot detect the FHSCs. After excluding the FHSCs,
which in fact constitute the majority
of VeLLOs according to our numerical simulations, it becomes evident that the 
numerically predicted number count of VeLLOs in the protostellar stage (0.68\%) is again much lower
than what is observationally inferred (6.25\%).
Finally, we note that 28 out of 31 hybrid accretion models (90\%) 
have total luminosities always exceeding the VeLLO upper limit. 


While Figure~\ref{fig4} provides the information about how
often and in what evolutionary phase we may expect to see
VeLLOs, it tells little about the physical properties of VeLLOs, such as their mass, for example.   
To determine the masses of VeLLOs, we present in 
Figure~\ref{fig5} the $L_{\rm tot}$--$M_{\rm obj}$
diagram for the 31 models of Table~\ref{table1}. The horizontal dashed lines
indicate the VeLLOs upper luminosity, $L_{\rm max}^{\rm vello}$.
The mosaic in the top-left panel shows the fraction of total time that all models spend
in various bins in the $L_{\rm obj}$--$M_{\rm obj}$ diagram. 
The red, orange, and blue dots  present the data for 
the FHSC, Class 0 and Class I phases, respectively.
Our modeling suggests that most VeLLOs are in fact brown dwarfs and/or very low-mass stars.
The maximum mass of the VeLLO is found to be approximately $0.12~M_\odot$.
Higher-mass stars are characterized by total luminosities that are
gradually increasing with mass and inevitably exceed the VeLLO's upper limit.

\subsubsection{The effect of variable accretion}
\label{varaccr}
Mass accretion rates in the embedded phase of star formation often 
demonstrate a highly variable character \citep[e.g.]{VB2010,VB2015,Elbakyan2016}.
Variable accretion with episodic bursts can, in particular, help to 
resolve the so-called luminosity problem \citep{DV2012}, whereby 
accretion luminosities typically observed for embedded protostars
are factors of 10--100 lower than predicted from the spherical
collapse models. In this section, we investigate to what extent
variable accretion has an effect on the population of VeLLOs in
the Class 0/I phases of star formation.

Many of our models also show variable accretion with episodic bursts, which 
is reflected in the corresponding variations of total luminosities 
(see Figure~\ref{fig1}). In order to artificially reduce the effect of 
variable accretion on the total luminosities, we apply the moving average to the total
luminosity in models 21-24 and 29-31 over a time period of 20~kyr and in the remaining models 
over a time period of 5~kyr. Longer time periods are needed to smooth strong 
luminosity variations in models with massive disks.
The resulting averaged luminosities are plotted with green solid 
lines in Figure~\ref{fig1}. The corresponding fractions of time spent
in the VeLLO state are
shown in Table~\ref{table3}. After averaging, the fraction of VeLLOs in the 
protostellar phase is further reduced. Now, the population of VeLLOs
is almost totally dominated by FHSCs, except probably for the Class I phase,
wherein we may still expect a rare occurrence of such objects.

\begin{table}
\renewcommand{\arraystretch}{2.0}
\begin{centering}
{\small{}}%
\begin{tabular}{|c|c|c|c|}
\multicolumn{1}{c}{} & \multicolumn{3}{c}{}\tabularnewline
\hline 
Phase & $\frac{t_{\rm VeLLOs}^{\rm phase}}{t_{\rm VeLLOs}^{\rm tot}}$ & 
$\frac{t_{\rm VeLLOs}^{\rm phase}}{t_{\rm phase}}$ & $\frac{t_{\rm VeLLOs}^{\rm phase}}{t_{\rm
tot}}$\tabularnewline
\hline 
{\small{}FHSC} & {\small{}97.0} & {\small{}100.0} & {\small{}6.23}\tabularnewline
\hline 
{\small{}Class 0} & {\small{}0.65} & {\small{}0.19} & {\small{}0.04}\tabularnewline
\hline 
{\small{}Class I} & {\small{}2.35} & {\small{}0.21} & {\small{}0.15}\tabularnewline
\hline 
\end{tabular}
\par\end{centering}{\small \par}

\protect\caption{\label{table3}Fraction of time (in per cent) spent by hybrid accretion models in each evolution  phase with total luminosities smoothed as described in Section~\ref{varaccr}. }
\end{table}

\subsection{Hot accretion}
\begin{figure*}
\begin{centering}
\includegraphics[scale=0.8]{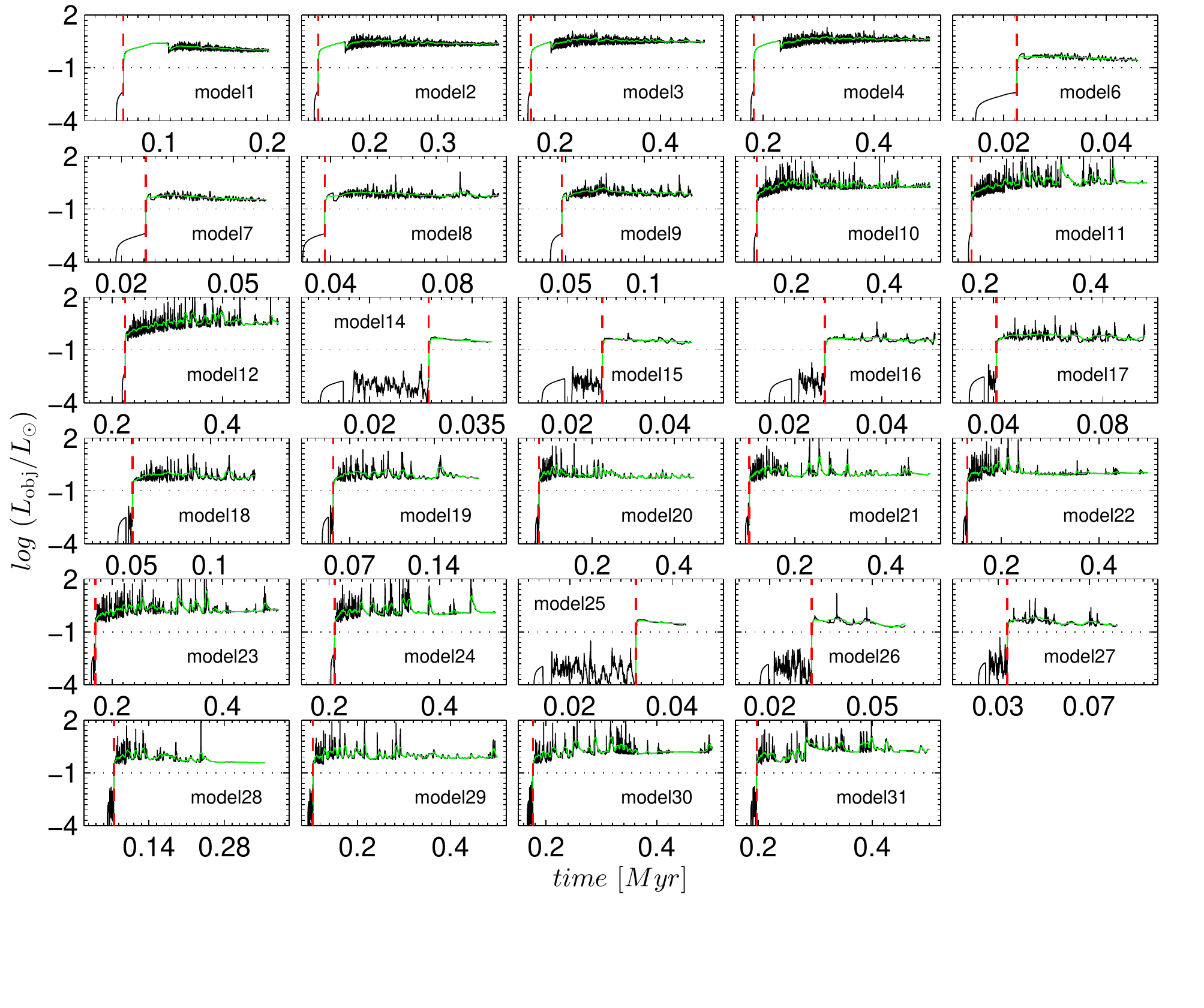}
\par\end{centering}
\protect\caption{Same as Figure~\ref{fig1}, but for 29 models with hot accretion.}
\label{fig5a}
\end{figure*}

 In this section, we consider the hot accretion scenario, in which $\alpha$ is always set to a constant
value. We have recalculated 29 models shown in Table~\ref{table1}, but with $\alpha=0.5$, irrespective of the actual mass accretion rate 
onto the star.  We note that two models were omitted due to numerical difficulties.
The resulting total luminosities as a function of time elapsed 
since the onset of numerical simulations are plotted in Figure~\ref{fig5a}.
For all considered models in the protostellar phase (to the right from the vertical red dashed line), $L_{\rm obj}$ are greater than $L_{\rm VeLLO}^{\rm max}=0.1~L_\odot$, 
effectively meaning no VeLLOs in the protostellar phase. 
This, however, contradicts the Spitzer observations \citep{Dunham2008} indicating that 
VeLLOs do exist in the protostellar stage, representing 6.25\% (7/112) of the total 
population of protostars (and perhaps more, as discussed in Section 5).

\begin{figure}
  \resizebox{\hsize}{!}{\includegraphics{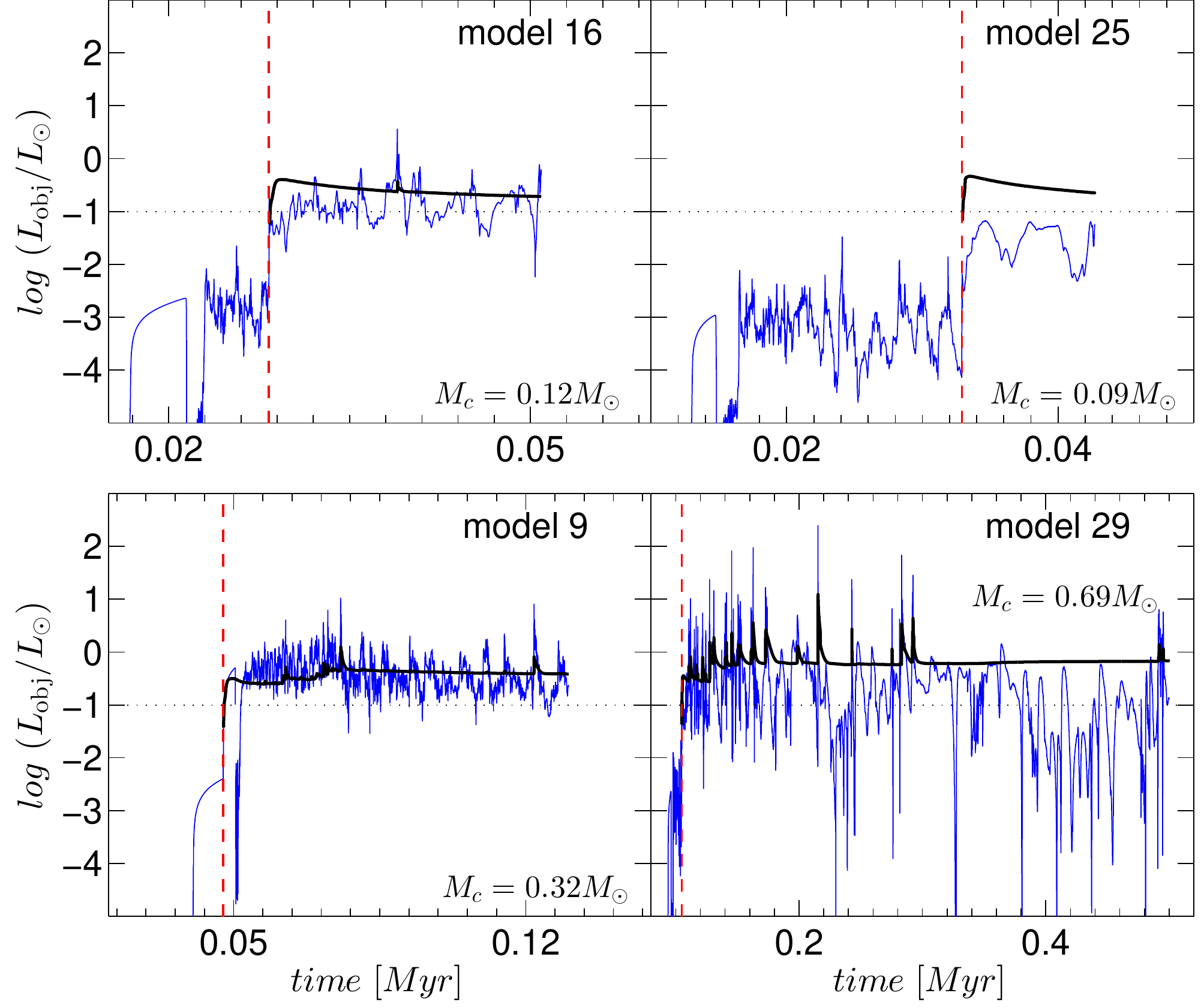}}
  \caption{Same as Figure~\ref{fig2}, but for hot accretion.}
  \label{fig5b}
\end{figure}

To further investigate this finding, in Figure~\ref{fig5b} we plot the accretion and photospheric luminosities
(rather than the sum of both) for the same four models as in Figure~\ref{fig2}. In all four hot 
accretion models, the photospheric luminosity (defining the floor for the total luminosity) 
is always higher than $L_{\rm VeLLO}^{\rm max}=0.1~L_\odot$, in contrast with the corresponding
hybrid accretion models, two of which having $L_{\rm phot}\ll L_{\rm VeLLO}^{\rm max}$. 
The same tendency is found for all other hot accretion models. The star bloats in response 
to the constant absorption of accretion energy, which acts to increase its photospheric luminosity.
We note that the accretion luminosities in the hot accretion
models are systematically lower than those in the hybrid ones due to the fact that half of the 
accreted energy is absorbed by the protostar\footnote{They also show a somewhat 
different pattern, which is
explained by the self-consistent nature of our coupled disk plus star numerical model, wherein stellar
output affects the disk evolution.}. This, however, does not help to produce VeLLOs, because the 
photospheric luminosity is too high.
To check how this finding depends on the adopted value of $\alpha$, we have recalculated  several models
with  $\alpha$ always set to 0.2 and found similar results.

\subsection{Cold accretion}

In this section, we present our results for the cold accretion scenario,
in which the fraction of accretion energy $\alpha$ absorbed by the protostar
is set to zero independent of the actual value of the mass accretion rate. 
The parameters of 32 cold accretion models are listed in
Table~\ref{table3a}. The cold accretion models differ somewhat from the 
hybrid accretion ones, because we had to introduce several of new models instead of
those that failed to run due to numerical reasons.

\begin{table}
\protect\caption{\label{table3a} Parameters for the
the cold accretion models}
\begin{centering}
\begin{tabular}{ccccc}
\hline 
Model & $M_{\mathrm{core}}$ & $\beta$  & $r_{\mathrm{out}}$ & $M_{*,\,\mathrm{fin}}$\tabularnewline
 & {[}$M_{\odot}${]} & {[}\%{]} & {[}AU{]} & {[}$M_{\odot}${]}\tabularnewline
\hline 
1 & 0.46 & 0.14 & 0.03 & 0.34\tabularnewline
2 & 0.62 & 0.14 & 0.04 & 0.36\tabularnewline
3 & 0.92 & 0.14 & 0.06 & 0.57\tabularnewline
4 & 1.38 & 0.14 & 0.09 & 0.79\tabularnewline
5 & 1.69 & 0.14 & 0.11 & 0.83\tabularnewline
6 & 2.00 & 0.14 & 0.13 & 0.84\tabularnewline
7 & 0.09 & 0.57 & 0.006 & 0.06\tabularnewline
8 & 0.12 & 0.57 & 0.008 & 0.07\tabularnewline
9 & 0.15 & 0.57 & 0.01 & 0.09\tabularnewline
10 & 0.31 & 0.57 & 0.02 & 0.16\tabularnewline
11 & 0.62 & 0.57 & 0.04 & 0.31\tabularnewline
12 & 0.77 & 0.57 & 0.05 & 0.39\tabularnewline
13 & 1.08 & 0.57 & 0.07 & 0.55\tabularnewline
14 & 1.69 & 0.57 & 0.11 & 0.77\tabularnewline
15 & 0.09 & 1.74 & 0.006 & 0.05\tabularnewline
16 & 0.11 & 1.74 & 0.007 & 0.054\tabularnewline
17 & 0.12 & 1.74 & 0.008 & 0.063\tabularnewline
18 & 0.15 & 1.74 & 0.01 & 0.08\tabularnewline
19 & 0.31 & 1.74 & 0.02 & 0.16\tabularnewline
20 & 0.46 & 1.74 & 0.03 & 0.23\tabularnewline
21 & 0.62 & 1.74 & 0.04 & 0.32\tabularnewline
22 & 0.77 & 1.27 & 0.05 & 0.34\tabularnewline
23 & 1.08 & 1.74 & 0.07 & 0.35\tabularnewline
24 & 1.38 & 1.74 & 0.09 & 0.34\tabularnewline
25 & 1.69 & 1.74 & 0.11 & 0.52\tabularnewline
26 & 0.09 & 2.87 & 0.006 & 0.058\tabularnewline
27 & 0.11 & 2.87 & 0.007 & 0.051\tabularnewline
28 & 0.14 & 2.87 & 0.009 & 0.067\tabularnewline
29 & 0.15 & 2.87 & 0.01 & 0.073\tabularnewline
30 & 0.46 & 2.87 & 0.03 & 0.25\tabularnewline
31 & 0.92 & 2.87 & 0.06 & 0.46\tabularnewline
32 & 1.23 & 2.87 & 0.08 & 0.49\tabularnewline
\hline 
\end{tabular}
\par\end{centering}

\end{table}

\begin{figure*}
\begin{centering}
\includegraphics[scale=0.8]{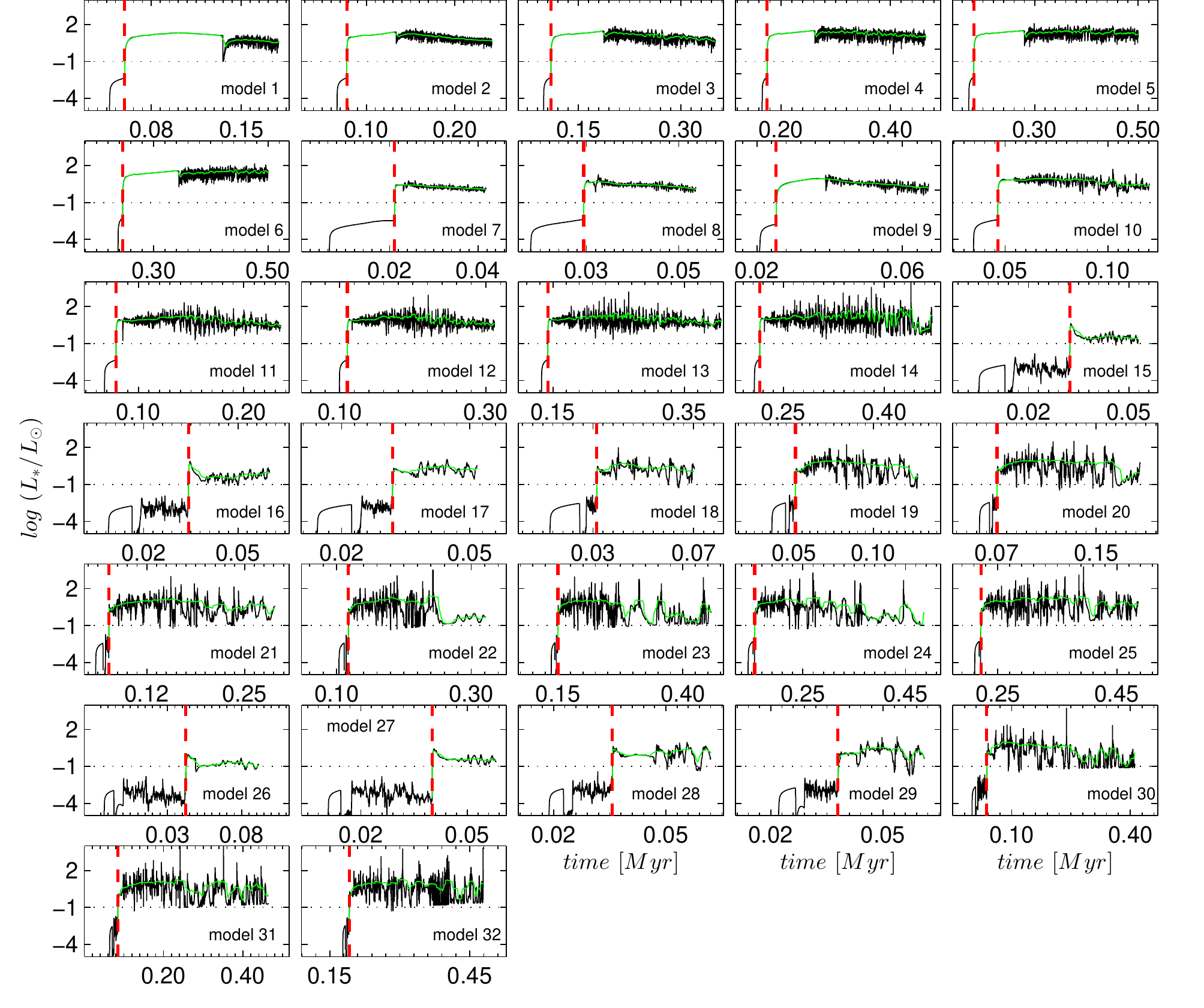}
\par\end{centering}
\protect\caption{\label{fig6} Similar to Figure~\ref{fig1}, but for 32 models with cold accretion.}
\end{figure*}

We start by examining the time evolution of the total luminosities $L_{\rm obj}$ 
shown in Figure~\ref{fig6} by the black solid lines. The vertical red dashed line
indicates the time instance of the formation of the protostellar/proto-brown dwarf
seed from the FHSC. The early FHSC phase is not affected by the choice of $\alpha$ ;
the total luminosity in this phase is always below the upper VeLLO limit of $0.1~L_\odot$.
In the protostellar phase, however, a certain difference between the hybrid and cold
accretion becomes evident. The variability amplitude in $L_{\rm obj}$ is notably
higher and  the minimum value in $L_{\rm obj}$ 
is notably lower in the cold accretion scenario than in the hybrid accretion one. 
As a consequence, more cold accretion models now
pass through the VeLLO state in the protostellar phase or at least come very close to the VeLLO state. 

\begin{figure}
  \resizebox{\hsize}{!}{\includegraphics{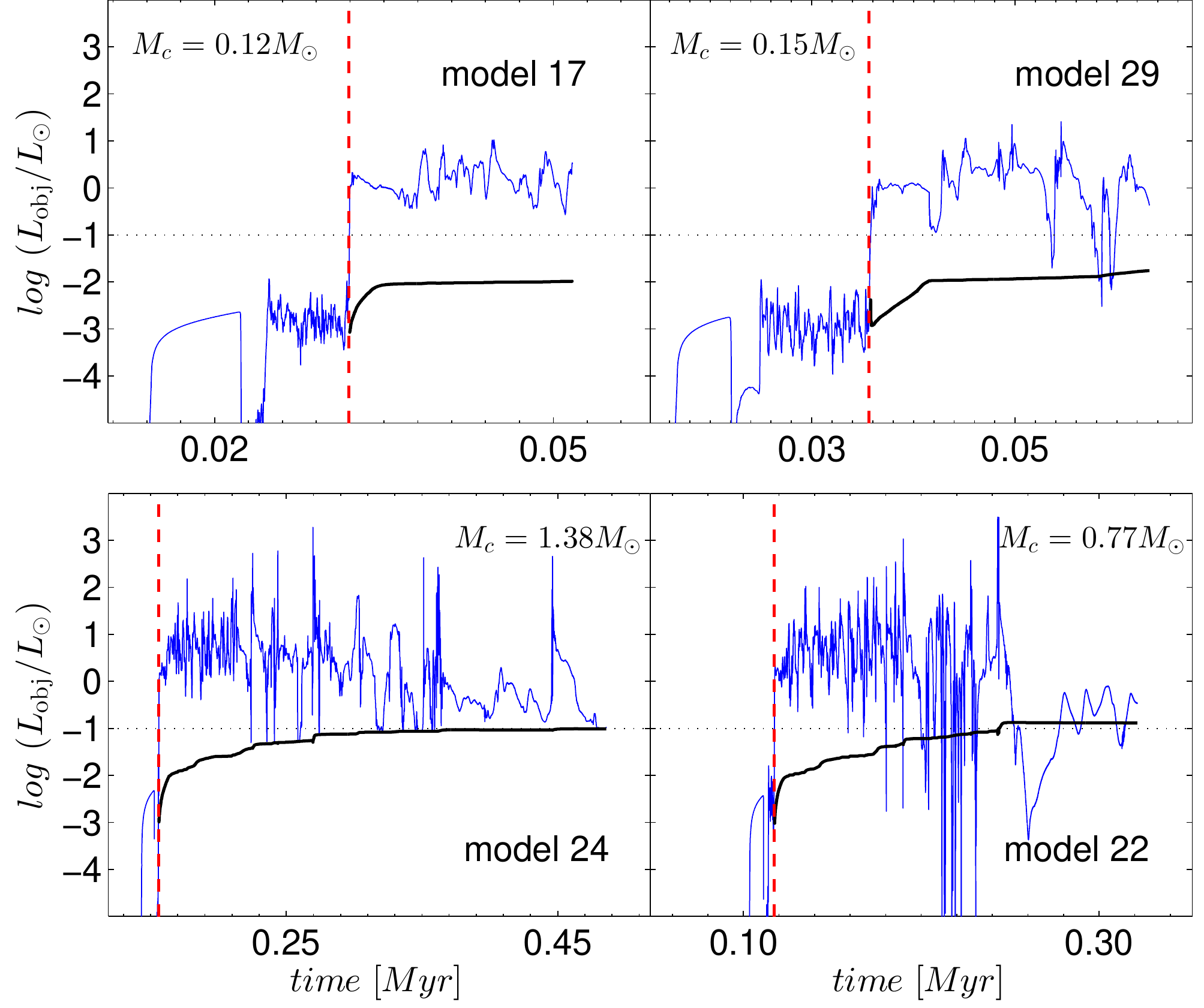}}
  \caption{Similar to Figure~\ref{fig2}, but for models with cold accretion.}
  \label{fig7}
\end{figure}

To illustrate this tendency, we plot in Figure~\ref{fig7} the accretion and photospheric  
luminosities vs. time for several prototype models with cold accretion. 
In particular, the thin blue and thick black lines show
the accretion and photospheric luminosities, respectively, while the vertical red 
dashed line separates 
the FHSC and protostellar phases. The horizontal dotted lines mark the 
maximum luminosity of VeLLOs, $L^{\rm vello}_{\rm max}=0.1~L_\odot$.
Contrary to the hybrid accretion models shown in Figure~\ref{fig2}, the cold accretion
models are often characterized by photospheric luminosity that is mostly lower than
the VeLLOs upper limit of 0.1~$L_\odot$, even for models with higher-$M_{\rm c}$ and $\beta$
(e.g., models 22 and 24).
This is because no accretion energy is absorbed by the protostar 
and its radius increases with time much slower than in the hybrid accretion scenario; 
see figure~4 in \citet{Baraffe12}. Smaller stellar radius means higher accretion luminosity
for similar accretion rates and stellar masses. In addition,
the photospheric luminosity sets the floor
for the total luminosity and  this floor value is lower in the cold accretion models. 
As a result, when variable accretion luminosity is added
to the (smooth) photospheric luminosity,  the cold accretion models show higher 
variability than the hybrid accretion models.

\begin{figure}
  \resizebox{\hsize}{!}{\includegraphics{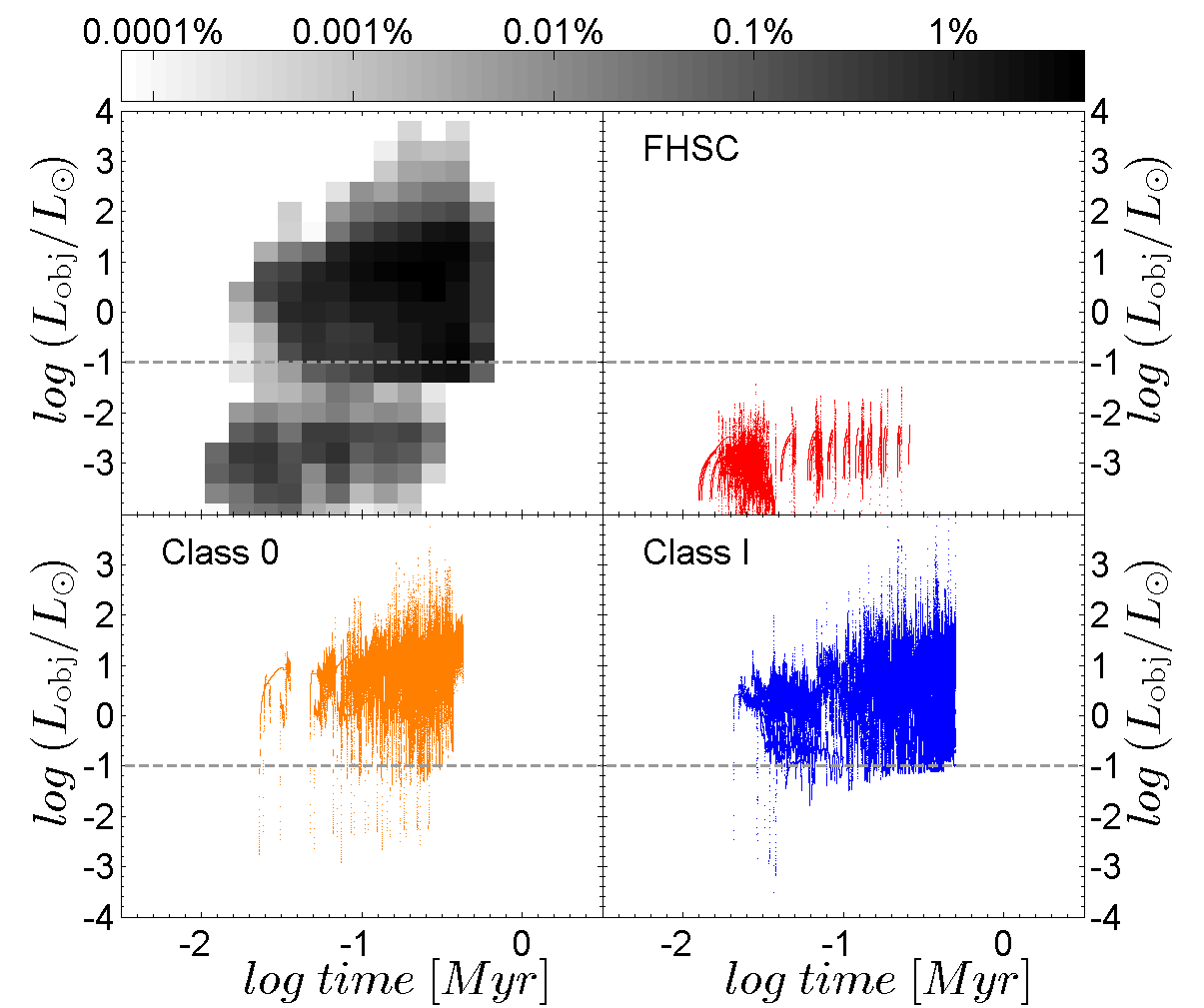}}
  \caption{Similar to Figure~\ref{fig4}, but for models with cold accretion.}
  \label{fig8}
\end{figure}

Following the method described in Section~\ref{statHA},
we calculate the fraction of total time $f_{\rm bin}$ that all cold accretion models spend
in various bins in the $L_{\rm obj}$--$t_{\rm evol}$ diagram. 
The resulting time fractions are shown by the gray-scale mosaic in the top-left panel of Figure~\ref{fig8}.
The red, orange, and blue dots present the model data for 
the FHSC, Class 0 and Class I phases, respectively. 
A visual comparison of Figures~\ref{fig4} and \ref{fig8} reveals that protostars with cold accretion
spend more time in the VeLLO state than that with hybrid accretion.
For instance, Class~I objects with cold accretion can be found in the VeLLO state
up to an age of $10^{-0.2}=0.65$~Myr, whereas the corresponding objects with hybrid accretion  
have total luminosities that always exceed the VeLLO's upper limit of 0.1~$L_\odot$ 
already after $10^{-0.75}=0.18$~Myr. 

The time fractions that the cold accretion models spend in the VeLLO state are summarized in 
Table~\ref{table4}. From comparison of Tables~\ref{table2} and \ref{table4} it becomes evident
that the cold accretion scenario predicts more VeLLOs in the protostellar phase than does the
hybrid accretion one. For instance, for the FHSCs, 
the combined fraction of time spent by the hybrid models in the VeLLO state 
(normalized to the total time spent in the VeLLOs state for all 
three phases) is 90.6\%, while for the cold accretion models the corresponding value is only 47\%.
A larger fraction of time in the VeLLO state is now spent in the protostellar phase: 1.76\% for the
Class~0 phase and 51.2\% for the Class~I phase. 
When considering the fractions of time spent in the VeLLO state normalized to the total 
evolution time $t^{\rm tot}$,
we see a similar tendency; the fraction of VeLLOs in the Class~I phase (5.62\%) 
slightly  exceeds the corresponding value for the FHSC phase (5.16\%).
When comparing the fraction of VeLLOs relative to the total duration of the embedded phase (last
column in Table~\ref{table4}), we see that the cold accretion scenario 
predicts most VeLLOs to actually occur in the Class I phase (5.62\%), contrary to the 
hybrid accretion scenario wherein most VeLLOs were seen in the FHSC phase. 
Interestingly, the predicted fraction of VeLLOs in the Class 0 is much smaller (0.19\%) than
in the Class I phase (5.62\%).
This seems to contradict the seemingly very young age of observed VeLLOs with 73\% (11/15) 
being Class 0 objects and only 27\% (4/15) being Class I ones \citep{Dunham2008}.
The existing VeLLO surveys may, however, be biased towards Class 0 sources as discussed
in Section~\ref{conclude}. Finally, we note that 16 out of 31 cold accretion models (50\%) 
have total luminosities always exceeding the VeLLO upper limit.

\begin{table}
\renewcommand{\arraystretch}{2.0}
\begin{centering}
{\small{}}%
\begin{tabular}{|c|c|c|c|}
\multicolumn{1}{c}{} & \multicolumn{3}{c}{}\tabularnewline
\hline 
Phase & $\frac{t_{\rm VeLLOs}^{\rm phase}}{t_{\rm VeLLOs}^{\rm tot}}$ & $\frac{t_{\rm VeLLOs}^{\rm
phase}}{t_{\rm phase}}$ & $\frac{t_{\rm VeLLOs}^{\rm phase}}{t_{\rm tot}}$\tabularnewline
\hline 
{\small{}FHSC} & {\small{}47.03 } & {\small{}100.0} & {\small{}5.16}\tabularnewline
\hline 
{\small{}Class 0} & {\small{}1.76} & {\small{}0.82} & {\small{}0.19}\tabularnewline
\hline 
{\small{}Class I} & {\small{}51.21 } & {\small{}7.91} & {\small{}5.62}\tabularnewline
\hline 
\end{tabular}
\par\end{centering}{\small \par}

\protect\caption{\label{table4}Fraction of time (in per cent) spent by cold accretion models in each evolution phase.}
\end{table}

\begin{figure}
  \resizebox{\hsize}{!}{\includegraphics{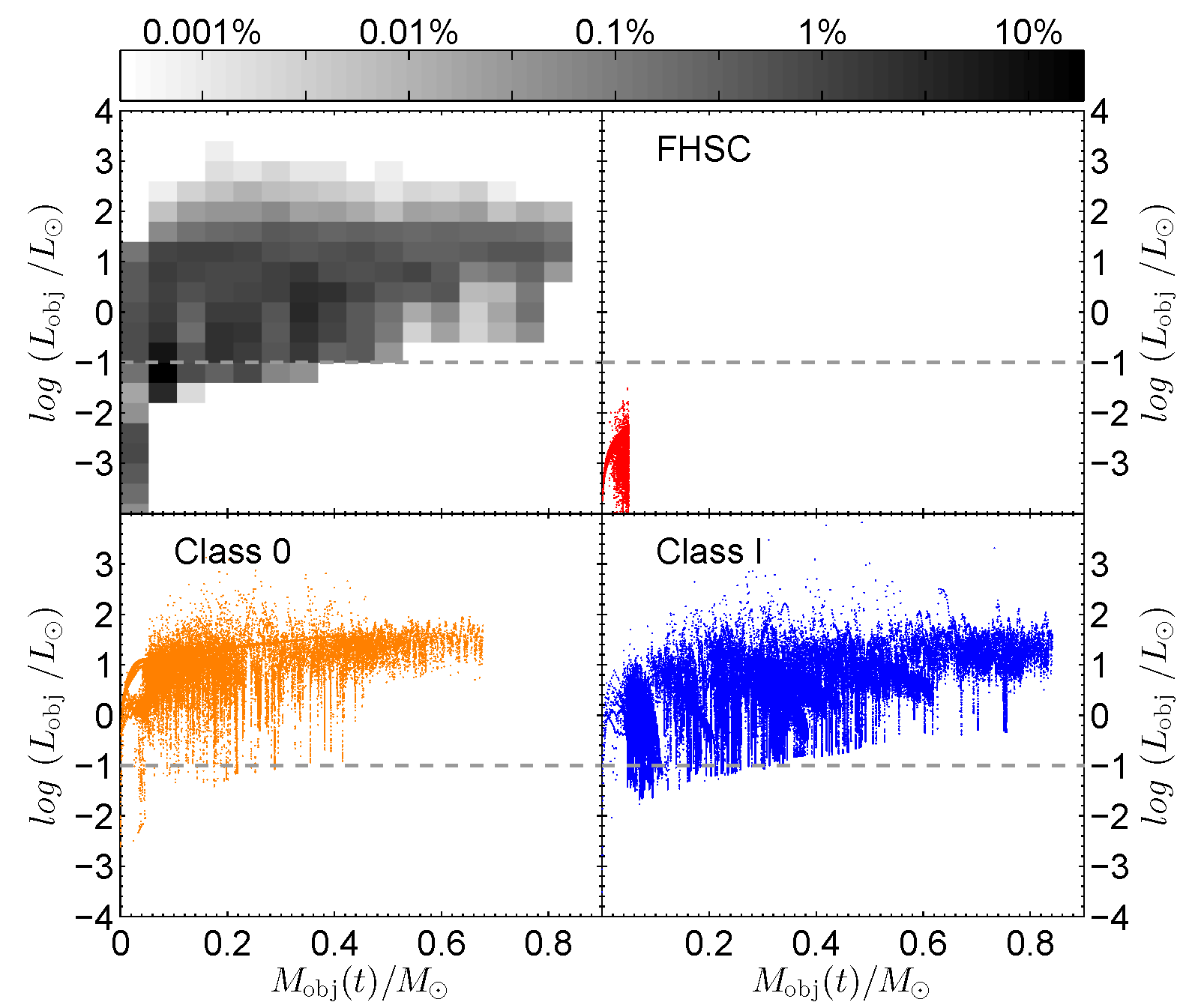}}
  \caption{Similar to Figure~\ref{fig5}, but for models with cold accretion.}
  \label{fig9}
\end{figure}

Figure~\ref{fig9} presents the total luminosity ($L_{\rm obj}$) vs. central object mass ($M_{\rm obj}$) diagram for 32 cold accretion models. A comparison of the $L_{\rm obj}$--$M_{\rm obj}$ diagram in 
Figure~\ref{fig9}
with that for the hybrid accretion models (Figure~\ref{fig5}) indicates that VeLLOs in the
cold accretion scenario can have a significantly higher protostellar mass than in the hybrid accretion
scenario. In the cold accretion scenario, the maximum mass of the protostar in the VeLLO state (for
both Class 0 and Class I phases)
is approximately $0.3~M_\odot$; at least a factor of 2 higher than in the hybrid accretion scenario.

Finally, we consider how variable accretion can affect the statistics of VeLLOs in the cold accretion
scenario. The green solid lines in Figure~\ref{fig6} show the total luminosities that are 
smoothed to reduce their time variability. We applied the same running average as was applied to the
hybrid accretion models in Section~\ref{varaccr}. The resulting fractions of time spent
in the VeLLO state are
shown Table~\ref{table5}. After averaging, the fraction of VeLLOs in the 
protostellar Class 0/I phases has decreased significantly. Similarly to the hybrid accretion case, 
the population of VeLLOs is now totally dominated by FHSCs, except probably for the Class 0 phase,
wherein we may still occasionally see such objects. We therefore conclude that accretion 
variability with
episodic bursts holds the key to the possible frequent occurrence of VeLLOs in the protostellar phase.

\begin{table}
\renewcommand{\arraystretch}{2.0}
\begin{centering}
{\small{}}%
\begin{tabular}{|c|c|c|c|}
\multicolumn{1}{c}{} & \multicolumn{3}{c}{}\tabularnewline
\hline 
Phase & $\frac{t_{\rm VeLLOs}^{\rm phase}}{t_{\rm VeLLOs}^{\rm tot}}$ & 
$\frac{t_{\rm VeLLOs}^{\rm phase}}{t_{\rm phase}}$ & $\frac{t_{\rm VeLLOs}^{\rm phase}}{t_{\rm
tot}}$\tabularnewline
\hline 
{\small{}FHSC} & {\small{}94.6} & {\small{}100.0} & {\small{}5.73}\tabularnewline
\hline 
{\small{}Class 0} & {\small{}0.41} & {\small{}0.1} & {\small{}0.02}\tabularnewline
\hline 
{\small{}Class I} & {\small{}5.0} & {\small{}0.43} & {\small{}0.3}\tabularnewline
\hline 
\end{tabular}
\par\end{centering}{\small \par}

\protect\caption{\label{table5}Similar to Table \ref{table3}, but for cold accretion}
\end{table}

\section{Discussion and model caveats}
\label{caveats}

To the best of our knowledge, this work is the first systematic attempt to
determine the nature of VeLLOs using numerical simulations of
coupled star plus disk evolution starting from the gravitational collapse
of pre-stellar cores. We also want to mention the works of \citet{Tomida2010} and \citet{Commercon2012},
who found that FHSCs are consistent with VeLLOs using post-processing with 
radiative transfer codes of 3D MHD calculations. 
Our numerical model, however, has limitations,  
which may to some extent affect our conclusions. 
We discuss the limitations in some detail below.

\subsection{ Initial conditions in pre-stellar cores} Our model cores 
were assumed to have a uniform and constant initial temperature of
10~K. However, an increase in the core temperature, for instance
due to the presence of nearby luminous massive stars, may have a twofold
effect on our results. First, this would produce
an increase in the mass infall rate, $\dot{M}_{\rm infall}$, 
onto the central object in the pre-disk phase or onto the disk in the post-disk phase,
because $\dot{M}_{\rm infall}$ is proportional to the cube of the sound speed.
For instance, a core temperature of 30~K would imply a five times
higher $\dot{M}_{\rm infall}$,  increasing the  accretion luminosity 
of FHSCs by the same fraction. Nevertheless, most FHSCs would still have
the accretion luminosity lower than the VeLLO upper limit of
$0.1~L_\odot$ (see Figures~\ref{fig4} and \ref{fig8}).
Second, a higher background temperature would act to weaken 
(however, not entirely suppressing) the gravitational instability in circumstellar disks
in the early stages of star formation. For instance, a background
temperature of 30~K reduces the accretion rate variability \citep{VB2010,VB2015},
which would also reduce the likelihood of a young protostar to enter the VeLLO state 
during quiescent, low-accretion-rate phases.

Furthermore, our model cores have the initial gas surface density distribution similar
to that of the integrated Bonnor-Ebert sphere with a small initial perturbation of $A=1.2$.
If the initial surface density perturbation is much higher 
(as might be expected for the shock compressed cores), then the mass infall rate
would also increase, simultaneously raising the accretion luminosity of the FHSCs.
According to radiation transfer simulations of \citet{Omukai2007}, the total luminosity
of the FHSC in the Shu spherical collapse model, wherein $\dot{M}_{\rm infall}=0.975 c_{\rm s}^{3}/G$,
varies in the $10^{-4}-10^{-3}~L_\odot$ limits, whereas in the Larson-Penston model, in which
$\dot{M}_{\rm infall}=46.9 c_{\rm s}^{3}/G$, the total luminosity is $0.01-0.1~L_\odot$.
Therefore, variations in the initial density perturbation $A$ may decrease the fraction of 
FHSCs in the VeLLO state. At the same time, a higher $\dot{M}_{\rm infall}$ in the post-disk formation
stage would increase the strength of accretion variability and may subsequently
increase the
fraction of VeLLOs in the protostellar Class 0/I phases. The effect of varying initial
conditions therefore requires further investigation.

\subsection{ The lifetime of the FHSCs} As summarized in \citet{Dunham2014}, the lifetime of the FHSCs
may vary in wide limits,  $0.5-50$~kyr, depending on the amount of rotation, the mass infall rate onto
the FHSC, and the strength of magnetic field in pre-stellar cores \citep[see also][]{Omukai2007,Tomida2010,Commercon2012}.
For instance, in the spherical collapse model of Shu, the lifetime of the FHSC is $3.2\times 10^4$~yr,
assuming the gas temperature of 10~K.
In the present work, we employed a toy model based on the fixed upper mass of the FHSC and 
the maximum gas temperature in the center of the core (see Section~\ref{numeric}), which yielded 
lifetimes of between $2.8\times 10^3$~yr and $2.2\times 10^4$~yr. These values span a somewhat 
narrower range than those obtained from more realistic simulations and the construction
of a more realistic model for the FHSC is needed for future work.

\subsection{ Internal disk luminosity} VeLLOs are observationally identified as objects with 
an internal luminosity $\le 0.1~L_\odot$, the latter including the contribution from both
the central object and circumstellar disk (if present). 
In our modeling, we calculated the total luminosity of the central object only,
and neglected the internal luminosity of the disk, which may be non-zero depending
on the rate of viscous and compressional heating in the disk. We currently develop 
disk models that include an accurate radiative transfer through the vertical disk 
column to estimate the internal disk luminosity.

\subsection{ The effect of the envelope}
The envelope reprocesses the stellar radiation, which reduces the emergent 
flux densities at shorter wavelengths and enhances them at longer ones. 
Our spectral energy distributions are sufficiently well-sampled and, to first order, luminosity is 
conserved since the loss of the short-wavelength radiation is compensated by the increase in 
the long-wavelength radiation \citep{enoch2009:protostars,Dunham2008,Dunham2013,Evans2009}.  
However, the second order effects, such as geometry of the outflow cavity 
and external heating, can modify the emergent flux. For instance,  the emergent
bolometric luminosities are, on average, reduced by 20\% relative
to the internal luminosity \citep{Frimann2016} when the envelope is not spherical, but feature outflows. External heating can add from $\ll 0.1 L_\odot$ up to several tenths of a solar luminosity (e.g., Evans et al. 2001), depending on the local strength 
of the interstellar radiation field, how embedded a core is within its parent cloud, and the core mass.
These effects were taken into account to the best of our ability,  and when we compare the 
numbers and fractions of VeLLOs in our models to observations, we do that directly to VeLLOs as identified via internal luminosity in \citet{Dunham2008}.

\subsection{ The fraction of ejected mass} In our models, we have assumed that
10\% of the mass that passes through the sink cell is subsequently ejected by 
jets/outflows. The actual value is, however, a poorly constrained parameter and may vary in 
wide limits from 0.01\% to 50\% \citep{Pelletier1992, Wardle1993,Shu1994}. 
This may have a certain effect on our results. For instance, a value
systematically
higher than 10\% of ejected mass would lead to a lower stellar mass and, simultaneously, a higher disk-to-star mass ratio.
Lower stellar masses imply lower photospheric luminosities (under other equal conditions) and higher
disk-to-star mass ratios imply stronger gravitational instability and, consequently, higher 
accretion variability.
Both may act to increase the number of VeLLOs in the protostellar phase and focused studies
are needed to assess the effect of ejected mass on the properties of stars and disks in general.

\subsection{ VeLLO fractions vs. initial core mass} 
\begin{figure}
  \resizebox{\hsize}{!}{\includegraphics{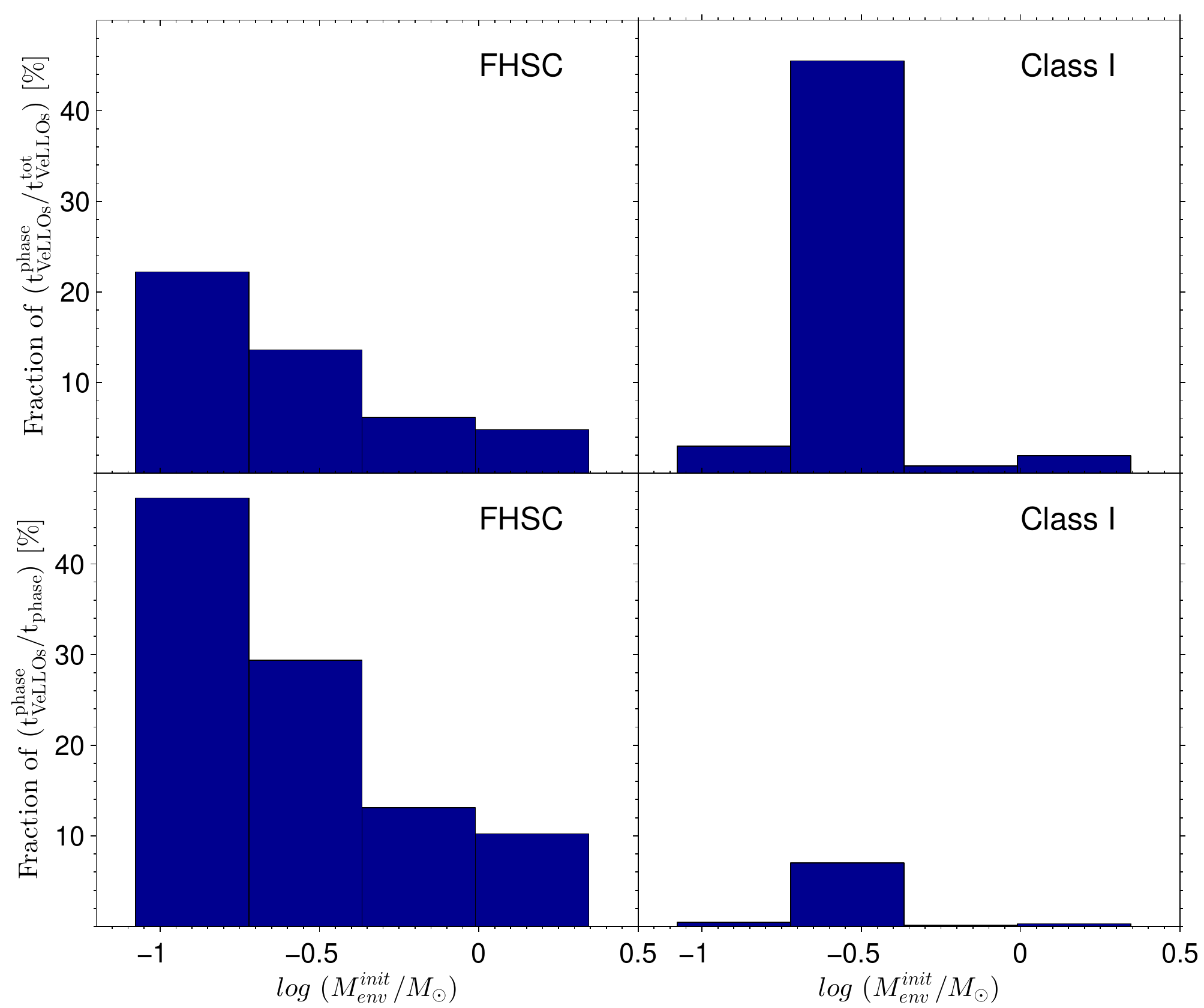}}
  \caption{Fraction of time spent by the cold accretion
models in the VeLLO state as a function of the initial core mass. }
  \label{fig12}
\end{figure}

It is interesting to know 
how the derived fractions of VeLLOs are related to the initial core mass in our models.
Figure~\ref{fig12} presents the histograms showing the fraction of time spent by the cold accretion
models in the VeLLO state as a function of the initial core mass (the latter represented by four 
mass bins). The sum of the fractions in each panel is equal to the total fraction of time 
spent by the corresponding objects (FHSC or Class I) in the VeLLO state. These total fractions
are shown in the first and second numerical columns of Table~\ref{table4}.  
We exclude from this analysis the Class 0 objects in the cold accretion scenario 
because of the low statistics. We do not consider the hybrid accretion models for the same reason.

Evidently, most VeLLOs in the FHSC phase are produced by low-mass cores, which
is merely a reflection of the slope of the IMF (see Fig.~\ref{fig3}). 
We note that the longevity 
of FHSCs may also depend on the initial rotation rate of the contracting 
pre-stellar cores \citep{Tomida2010}, which is a poorly constrained
parameter. If, for instance, lower-mass cores have systematically higher 
ratios of rotational to gravitational energy, than they may live longer and 
this may further strengthen the found tendency.
For the case of Class I objects, most VeLLOs are produced by cores with mass $M_{\rm c}\sim 0.3-0.4~M_\odot$.
Interestingly, this is approximately the minimum mass required for contracting cores to 
produce gravitationally (and fragmentationally) unstable disks \citep{Vorobyov2013}.
These systems will exhibit variable accretion with episodic bursts; a necessary prerequisite 
for VeLLOs in the protostellar stage according to our findings. At the same time, 
central objects produced by these cores will have systematically lower photospheric luminosities
(sometimes lower than $0.1~L_\odot$) than protostars produced by higher-mass cores (see Figure~\ref{fig7}).
This combination of effects enables more VeLLOs in the $M_{\rm c}=0.3-0.4~M_\odot$ mass range. 

Observationally constrained core masses for VeLLOs are known only for a handful of objects.
These are L1014 with $M_{\rm core}=1.7~M_\odot$ \citep{young2004:l1014}, 
IRAM04191 with  $M_{\rm core}=2.5~M_\odot$ \citep{Dunham2006}, L1521F
with $M_{\rm core}=4.8~M_\odot$ \citep{bourke2006:l1521f}, L328 with 
$M_{\rm core}=0.07~M_\odot$ \citep{lee2009:l328}, and 
L673-7 with $M_{\rm core}=2.0~M_\odot$ \citep{Dunham2010a}.
We note that these are current core masses and the initial core masses must have
been somewhat higher. Evidently, most VeLLOs belong to rather high core masses, in disagreement
with our modeling. We should emphasize, however, that the statistics in both cases (modeling and
observations) is rather poor. More work is needed to understand the reason of this 
mismatch.

\subsection{ Constant star formation rate} When converting our numerically predicted
time fractions to number fractions in Section~\ref{statHA}, we made a fundamental
assumption of a constant star formation rate.
\citet{PS1999,PS2000} claimed that star formation in a series of 
star forming regions (e.g., in the Orion Nebula Cluster) shows accelerated
star formation rates within the past few Myr. In Taurus, star formation
shows this effect specifically in filaments while the star formation rate
is declining outside filaments, after a peak approximately 2-3~Myr ago 
\citep{PS2002}. Cores in different stages 
\citep[e.g.,][]{Tafalla2015}, Class 0 and I protostars and classical
T Tauri stars co-exist, even if at different, specific locations and with different 
spatial spreads. This observational evidence suggests to us that star formation 
in active molecular clouds is ongoing for at least a couple of Myr if considered 
globally over the entire star formation region.

However, in our work we are predominantly interested in the evolution of
the star formation rate {\it within} 1~Myr, the time scales from FHSC
to protostars. Ignoring substructures and aiming at statistics of objects
in molecular clouds as a whole, and assuming that cloud-to-cloud 
variations can be averaged out, we find it appropriate to assume, to first
order, constant star formation rate over time scales of a few 100 kyr.


\begin{table}
\renewcommand{\arraystretch}{2.0}
\begin{centering}
{\small{}}%
\begin{tabular}{|c|c|c|c|}
\multicolumn{1}{c}{} & \multicolumn{3}{c}{}\tabularnewline
\hline 
Phase & $\frac{t_{L<0.2L_{\odot}}^{\rm phase}}{t_{L<0.2L_{\odot}}^{\rm tot}}$ & $\frac{t_{L<0.2L_{\odot}}^{\rm
phase}}{t_{\rm phase}}$ & $\frac{t_{L<0.2L_{\odot}}^{\rm phase}}{t_{\rm tot}}$\tabularnewline
\hline 
{\small{}FHSC} & {\small{}76.2 } & {\small{}100.0} & {\small{}6.62}\tabularnewline
\hline 
{\small{}Class 0} & {\small{}1.21} & {\small{}0.48} & {\small{}0.11}\tabularnewline
\hline 
{\small{}Class I} & {\small{}22.05 } & {\small{}2.68} & {\small{}1.92}\tabularnewline
\hline 
\end{tabular}
\par\end{centering}{\small \par}

\protect\caption{\label{table6}Similar to Table \ref{table2}, but for the upper VeLLO luminosity of
0.2~$L_\odot$.}
\end{table}

\begin{table}
\renewcommand{\arraystretch}{2.0}
\begin{centering}
{\small{}}%
\begin{tabular}{|c|c|c|c|}
\multicolumn{1}{c}{} & \multicolumn{3}{c}{}\tabularnewline
\hline 
Phase & $\frac{t_{L<0.2L_{\odot}}^{\rm phase}}{t_{L<0.2L_{\odot}}^{\rm tot}}$ & $\frac{t_{L<0.2L_{\odot}}^{\rm
phase}}{t_{\rm phase}}$ & $\frac{t_{L<0.2L_{\odot}}^{\rm phase}}{t_{\rm tot}}$\tabularnewline
\hline 
{\small{}FHSC} & {\small{}26.68 } & {\small{}100.0} & {\small{}5.16}\tabularnewline
\hline 
{\small{}Class 0} & {\small{}2.09} & {\small{}1.53} & {\small{}0.36}\tabularnewline
\hline 
{\small{}Class I} & {\small{}68.22 } & {\small{}16.68} & {\small{}11.87}\tabularnewline
\hline 
\end{tabular}
\par\end{centering}{\small \par}

\protect\caption{\label{table7}Similar to Table \ref{table4}, but for the upper VeLLO luminosity of
0.2~$L_\odot$.}
\end{table}


\section{Conclusions}
\label{conclude}
In this work, we have numerically studied the simultaneous evolution of protostars and 
protostellar disks by means of numerical hydrodynamics simulations coupled with a stellar evolution
code. We consider the gravitational collapse of a large set of model cores, starting  
from the pre-stellar phase and terminating at the end of the
embedded phase when 90\% of the initial core mass has accreted onto the forming protostar plus disk
system. The formation of the first hydrostatic core (FHSC) and its evolution is treated semi-analytically.
Three protostellar accretion scenarios are considered: hybrid accretion,
in which a fraction of accreted energy absorbed by the protostar depends on the accretion rate,
 hot accretion, wherein a constant fraction of the accreted energy is always absorbed by the protostar
irrespective of the accretion rate, and cold accretion, wherein all accretion energy is radiated 
away, regardless of the actual accretion
rate.  We aim at determining the nature of VeLLOs, which
are defined by the total luminosity $L_{\rm obj}$ not exceeding an upper limit of 0.1~$L_\odot$.
We found the following.

In the hybrid accretion scenario, most VeLLOs are expected to be 
FHSCs and only a small fraction are true protostars (see Table~\ref{table2}). 
When comparing the fraction of time spent by our models 
in the VeLLO state ($L_{\rm obj}<0.1~L_\odot$), we found  
that FHSCs occupy 90.63\% of the total VeLLO time,
whereas Class 0 and I objects have only $0.69\%$ and $8.68\%$, respectively.
Furthermore, FHSCs spend all their time in the VeLLO state, whereas
Class 0 and Class I protostars spend only a very small
fraction of the corresponding phase duration, meaning
that it would be unlikely (from a statistical point of view)  
to detect Class 0 and Class I protostars in the VeLLO state. 
Finally, VeLLOs are relatively rare objects, 
occupying only $7.3\%$ of the total duration of the embedded phase (the sum of the last
numerical column in Table~\ref{table2}), and most belong to the FHSCs (6.62\%) 
and only a very small fraction are Class 0 (0.05\%) and Class I (0.65\%) protostars.


In the cold accretion scenario, the situation with VeLLOs
is somewhat different (see Table~\ref{table4}). The majority of VeLLOs belong now
to the Class I phase of stellar evolution.  As with hybrid accretion, 
FHSCs spend all their time in the VeLLO state, but
the Class I protostars now spend a modest 
fraction (7.91\%) of the corresponding phase duration in the VeLLO state,
which is much greater than a negligible value of $0.89\%$ for the hybrid accretion models. 
This difference from the hybrid accretion case is caused by the reduced 
photospheric luminosity (and, as a consequence, reduced total luminosity) 
due to smaller stellar radii in the cold accretion models \citep{Baraffe12}.
However, VeLLOs are still rather rare objects,
occupying only $\approx 11\%$ of the embedded phase duration (the sum
of the last numerical column is Table~\ref{table4}),  but now most belong
to the Class I protostars (5.62\%) and FHSCs (5.16\%), and only a small fraction
are Class 0 protostars (0.19\%).

In the hot accretion scenario, all VeLLOs belong to
the FHSCs. No VeLLOs are found in either Class 0 or Class I phases. This is explained
by the fact that the photospheric accretion of forming protostars is always 
higher than the VeLLO upper limit of $0.1~L_\odot$.
Protostars bloat in response 
to the constant absorption of accretion energy and their 
photospheric luminosity raises above the VeLLO limit.

The most comprehensive search for VeLLOs to date was published in \citet{Dunham2008}
based on the Spitzer Space Telescope Legacy Project `From Molecular
Cores to Planet Forming Disks' (c2d).
More recent protostar searches, summarized in \citet{Dunham2014}, 
focused on bolometric rather than internal luminosity.  Since bolometric luminosity 
includes external heating, and this term can dominate over internal when the internal luminosity 
is $<0.1 L_\odot$, the distinction is crucial for finding VeLLOs. In \citet{Dunham2008}, 
15 VeLLOs were found in the c2d dataset, which contains a total of 112 protostars \citep{Evans2009}. Excluding sources from the Dunham et al. work that were not included in the Evans et al. 
paper, as that  focused on the large molecular clouds only,  there are 7 VeLLOs out of 112 protostars, suggesting a total fraction of 6.25\% for VeLLOs in the protostellar stage.  
Here, we should note that all 
VeLLOs in the Dunham et al. sample belong to the protostellar stage, because the 
FHSCs are undetectable with the Spitzer observations at 3.6--8 $\mu$m.
The obtained number is much higher than the fraction of 0.68\% that we obtained 
in our numerical simulations for hybrid accretion, but similar to what was found for cold accretion;
5.7\% (last numerical columns in Tables~\ref{table2} and \ref{table4}, but only for the sum of Class
0 and Class I phases). Here we should note that the observationally derived
VeLLOs fraction of 6.25\% may be a lower limit, because the existing surveys likely miss some objects
in the tenuous Class I phase.  The VeLLO sample needs to be revisited as the Herschel and SCUBA-2 core catalogs are published, since these are much deeper than the previous generation of surveys used by \citep{Dunham2008} and subsequent papers.

Most VeLLOs in the FHSC phase are produced by low-mass cores, which
can be explained by the slope of the IMF. This tendency may change somewhat (strengthen or weaken) if
the lifetime of the FHSCs depends systematically on the core 
initial rotation rate \citep[e.g.,][]{Tomida2010}. 
In the protostellar stage, most VeLLOs are produced by cores with mass $M_{\rm c}\sim 0.3-0.4~M_\odot$,
which is in disagreement with available measurements of the core masses indicating that 
most VeLLOs have more massive cores. The low statistics of both numerical simulations and observations
does not allow us to make definite conclusions and the reason for this disagreement remains to be understood.

VeLLOs are expected to be low-mass objects. In the 
hybrid accretion models, the maximum VeLLO mass is found to be 0.12~$M_\odot$,
while in the cold accretion case it is greater by a factor of approximately 2.5.
For higher-mass protostars, the total luminosity in the embedded phase of star formation is 
inevitably greater than the VeLLO's upper limit of 0.1~$L_\odot$.
Our work provides results that are testable with future 
ALMA observations that use Keplerian rotation to measure VeLLO masses.  
Such observations are needed to make further progress regarding the nature of VeLLOs.

Accretion variability with episodic bursts,
inherent to many our models thanks to gravitational instability and fragmentation
\citep{VB2015}, has a profound effect on the calculated fractions. When
variations in total luminosity caused by accretion variability are artificially
reduced (by smoothing the luminosities), the fraction of VeLLOs  in the Class 0/I phases greatly
reduces, regardless of the accretion scenario. Without accretion variability, 
the population of VeLLOs would be almost totally dominated by FHSCs.
We emphasize that episodic bursts and accretion variability 
can also be caused by mechanisms other than gravitational instability 
\citep[see a review by][]{audard2014:ppvi}. It remains to be investigated how these
additional sources of variability can modify our findings and hopefully
improve the agreement of models with observations.

In studying the nature of VeLLOs, we have assumed that these objects have 
an upper luminosity of 0.1~$L_\odot$. This value, however, has no physical 
motivation and is simply a result of useful convention.  If the upper luminosity
of VeLLOs, for instance, is increased to 0.2~$L_\odot$, then the fraction of VeLLOs
in the protostellar phase, as well as the overall fraction of VeLLOs relative to
normal stars, also increases by a factor of approximately 3 for hybrid accretion and 
by a factor of 1.5-2.0 for cold accretion. The corresponding exact values are provided in Tables~\ref{table6}
and Table~\ref{table7}, respectively.

One interesting feature that can be
seen in Tables~\ref{table2} and \ref{table4} is that the fraction of VeLLOs in the Class~0
phase is systematically lower than in the later Class~I phase. This finding seems to contradict 
observations, suggesting more VeLLOs in the Class 0 phase with  73\% (11/15) 
being Class 0 objects and only 27\% (4/15) being Class I objects \citep{Dunham2008}.
This is likely caused by three factors.
Firstly, accretion variability is lower in the early Class 0 phase \citep[e.g.,][]{VB2015},
mainly due to lower disk masses and consequently reduced gravitational 
instability. Secondly, the duration of the Class 0 phase is shorter than that of the Class I phase,
particularly for low-mass cores, owing to the fact that the FHSCs need to accumulate 
$\approx 0.05~M_\odot$ for the second collapse to ensue, which may take a sizeable fraction of 
the total duration of the embedded phase. We note that this argument 
applies strictly 
to the fractions calculated relative to the total time of all phases $t_{\rm tot}$ and to the total
time spend in the VeLLO state $t^{\rm tot}_{\rm VeLLOs}$, but not to the time spend in each phase
$t_{\rm phase}$. Thirdly, on the 
observational side, \citet{Dunham2008} may be biased against Class~I 
sources (as discussed above) since they required evidence that 
the Spitzer sources were embedded in dense cores, and, for Class~I sources, the 
cores may be tenuous enough to avoid detection.

Finally, we note that if most VeLLOs are FHSCs, then large number detections of VeLLOs would 
imply long FHSC lifetimes.
If rotation is not taken into account, core collapse simulations predict short lifetimes for 
FHSCs, in the order of several thousand years \citep{Omukai2007}. Long lifetimes would imply that  
 rotation and/or magnetic fields play an important role in the evolution of these objects 
 \citep{Tomida2010}.

{\it Acknowledgments.}  The authors are thankful to the anonymous referee for useful comments
that helped to improve the paper. EIV is thankful to Isabelle Baraffe and Gilles Chabrier
for the Lyon stellar evolution code. This work is partly supported by the Austrian Science Fund (FWF) under research grant I2549-N27 and by the Russian Ministry of Education and Science Grant 3.961.2014/K. V.G.E. acknowledges Osterreichischer Austauschdienst (Austrian Agency for International Cooperation in Education and Research) for Ernst Mach grant.
The simulations were performed on the Vienna Scientific Cluster (VSC-2),
 on the Shared Hierarchical 
Academic Research Computing Network (SHARCNET), and on the Atlantic Computational Excellence Network 
(ACEnet). This publication is supported by the Austrian Science Fund (FWF).  
MMD acknowledges support from the Submillimeter Array (SMA) through a 
SMA postdoctoral fellowship, and from NASA through grant NNX13AE54G.

\end{document}